\documentclass[twocolumn,showpacs,preprintnumbers,amsmath,amssymb,nofootinbib,superscriptaddress]{revtex4} 
\usepackage{mathrsfs, epsfig, graphicx, url, dcolumn, bm}

\newcommand{\milliK}{{\rm mK}}

\newcommand{\Mpc}{{\rm Mpc}}
\newcommand{\perMpc}{\ensuremath{\Mpc^{-1}}}

\newcommand{\eV}{{\rm eV}}

\newcommand{\Ob}{\ensuremath{\Omega_{\rm b}}}

\newcommand{\Ok}{\ensuremath{\Omega_{\rm k}}}
\newcommand{\Ol}{\ensuremath{\Omega_\Lambda}}
\newcommand{\Om}{\ensuremath{\Omega_{\rm m}}}

\newcommand{\On}{\ensuremath{\Omega_\nu}}
\newcommand{\ob}{\ensuremath{\Omega_{\rm b} h^2}}

\newcommand{\om}{\ensuremath{\Omega_{\rm m} h^2}}
\newcommand{\on}{\ensuremath{\Omega_\nu h^2}}

\newcommand{\ns}{\ensuremath{n_{\rm s}}}
\newcommand{\nt}{\ensuremath{n_{\rm t}}}
\newcommand{\al}{\ensuremath{\alpha}}

\newcommand{\As}{\ensuremath{A_{\rm s}}}

\newcommand{\mnu}{\ensuremath{m_\nu}}
\newcommand{\xH}{\ensuremath{\bar{x}_{\rm H}}}
\newcommand{\xI}{\ensuremath{\bar{x}_{\rm i}}}
\newcommand{\Ts}{\ensuremath{T_S}}
\newcommand{\Tcmb}{\ensuremath{T_{\rm CMB}}}

\newcommand{\PDT}{\ensuremath{P_{\Delta T}}}
\newcommand{\Pxx}{\ensuremath{P_{\rm xx}}}
\newcommand{\Pxd}{\ensuremath{P_{{\rm x}\delta}}}
\newcommand{\Pdd}{\ensuremath{P_{\delta\delta}}}
\newcommand{\sPdd}{\ensuremath{\mathscr{P}_{\delta\delta}}}
\newcommand{\sPxd}{\ensuremath{\mathscr{P}_{x\delta}}}
\newcommand{\sPxx}{\ensuremath{\mathscr{P}_{xx}}}
\newcommand{\Pmuzero}{\ensuremath{P_{\mu^0}}}
\newcommand{\Pmutwo}{\ensuremath{P_{\mu^2}}}
\newcommand{\Pmufour}{\ensuremath{P_{\mu^4}}}
\newcommand{\Pmusix}{\ensuremath{P_{\mu^6}}}
\newcommand{\bsqxx}{\ensuremath{b^2_{\rm xx}}}
\newcommand{\Rxx}{\ensuremath{R_{\rm xx}}}
\newcommand{\alxx}{\ensuremath{\al_{\rm xx}}}
\newcommand{\gaxx}{\ensuremath{\gamma_{\rm xx}}}
\newcommand{\bsqxd}{\ensuremath{b^2_{{\rm x}\delta}}}
\newcommand{\Rxd}{\ensuremath{R_{{\rm x}\delta}}}
\newcommand{\alxd}{\ensuremath{\al_{{\rm x}\delta}}}

\newcommand{\kmax}{\ensuremath{k_{\rm max}}}
\newcommand{\kmin}{\ensuremath{k_{\rm min}}}

\newcommand{\onesig}{\ensuremath{1\sigma}}

\newcommand{\kpar}{\ensuremath{k_{\parallel}}}
\newcommand{\uperp}{\ensuremath{{\bf u}_{\perp}}}
\newcommand{\upar}{\ensuremath{u_{\parallel}}}
\newcommand{\bfk}{{\bf k}}

\newcommand{\bfx}{{\bf x}}
\newcommand{\bfu}{{\bf u}}
\newcommand{\bfV}{{\bf V}}
\newcommand{\bfy}{{\bf y}}
\newcommand{\bfF}{{\bf F}}

\newcommand{\etal}{{et al. }}
\newcommand{\ie}{{\frenchspacing i.e. }}
\newcommand{\eg}{{\frenchspacing e.g. }}

\newcommand{\ben}{\begin{equation}}
\newcommand{\een}{\end{equation}}
\newcommand{\bena}{\begin{eqnarray}}
\newcommand{\eena}{\end{eqnarray}}
\newcommand{\beq}[1]{\begin{equation}\label{#1}}
\newcommand{\eeq}{\end{equation}}
\newcommand{\Eq}[1]{Eq.~(\ref{#1})}
\newcommand{\eq}[1]{Eq.~\ref{#1}}
 
\newcommand{\Tab}[1]{Table~\ref{#1}}         

\newcommand{\bem}{\begin{enumerate}}
\newcommand{\eem}{\end{enumerate}}
\newcommand{\bit}{\begin{itemize}}
\newcommand{\eit}{\end{itemize}}

\newcommand{\la}{\lesssim}
\newcommand{\nodata}{\ensuremath{\cdots}}
\newcommand{\mnras}{Mon.~Not.~Roy.~Astron.~Soc. }

\begin{document}
\input{epsf.sty}

\title{How accurately can 21 cm tomography constrain cosmology?}

\author{Yi Mao}
\email{ymao@mit.edu}
\affiliation{Center for Theoretical Physics, Dept.~of Physics, Massachusetts Institute of Technology, Cambridge, MA 02139, USA} 
\author{Max Tegmark}
\email{tegmark@mit.edu}
\affiliation{Center for Theoretical Physics, Dept.~of Physics, Massachusetts Institute of Technology, Cambridge, MA 02139, USA}
\affiliation{MIT Kavli Institute for Astrophysics and Space Research, Cambridge, MA 02139, USA}
\author{Matthew McQuinn}
\affiliation{Harvard-Smithsonian Center for Astrophysics, 60 Garden Street, Cambridge, MA 02138, USA} 
\author{Matias Zaldarriaga}
\affiliation{Harvard-Smithsonian Center for Astrophysics, 60 Garden Street, Cambridge, MA 02138, USA} 
\affiliation{Jefferson Laboratory of Physics, Harvard University, Cambridge, MA 02138, USA}
\author{Oliver Zahn}
\affiliation{Harvard-Smithsonian Center for Astrophysics, 60 Garden Street, Cambridge, MA 02138, USA} 
\affiliation{Berkeley Center for Cosmological Physics, Department of Physics, University of California, and Lawrence Berkeley National Labs, 1 Cyclotron Road, Berkeley, CA 94720, USA}

\date{Submitted to Phys.~Rev.~D. February 21 2008; Accepted April 29 2008; Published July 25 2008}

\begin{abstract}

There is growing interest in using 3-dimensional neutral hydrogen
mapping with the redshifted 21 cm line as a cosmological probe.
However, its utility depends on many assumptions. To aid experimental
planning and design, we quantify how the precision with which
cosmological parameters can be measured depends on a broad range of
assumptions, focusing on the 21 cm signal from $6 < z < 20$. We cover
assumptions related to modeling of the ionization power spectrum,
to the experimental specifications like array layout and detector
noise, to uncertainties in the reionization history,
and to the level of contamination from astrophysical foregrounds. 
We derive simple analytic estimates for how various assumptions affect
an experiment's sensitivity, and we find that the modeling of
reionization is the most important, followed by the array layout.  We
present an accurate yet robust method for measuring cosmological
parameters that exploits the fact that the ionization power spectra
are rather smooth functions that can be accurately fit by $7$
phenomenological parameters.
We find that for future experiments, marginalizing over these nuisance
parameters may provide almost as tight constraints on the cosmology as
if 21 cm tomography measured the matter power spectrum directly. A
future square kilometer array optimized for 21 cm tomography could
improve the sensitivity to spatial curvature and neutrino masses by up
to two orders of magnitude, to $\Delta\Omega_k\approx 0.0002$ and
$\Delta m_\nu\approx 0.007$ eV, and give a $4\sigma$ detection of the
spectral index running predicted by the simplest inflation models.

\end{abstract}


\maketitle

\section{Introduction}
\label{sec:intro}

\begin{figure}[h!]
\vskip-0.5cm
\includegraphics[width=0.5\textwidth]{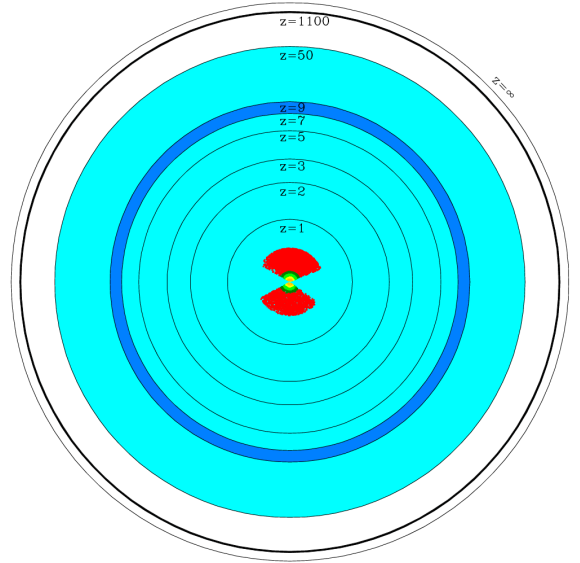}
\vskip-1.2cm
\caption{ 21 cm tomography can potentially map most of our observable
universe (light blue/light grey), whereas the CMB probes mainly a thin shell
at $z\sim 10^3$ and current large-scale structure surveys (here
exemplified by the Sloan Digital Sky Survey and its luminous red
galaxies) map only small volumes near the center.  This paper focuses
on the convenient $7\la z \la 9$ region (dark blue/dark grey).
\label{fig:spheres}
}
\end{figure}

Three-dimensional mapping of our Universe using the redshifted 21 cm
hydrogen line has recently emerged as a promising cosmological probe,
with arguably greater long-term potential than the cosmic microwave
background (CMB).  The information garnered about cosmological
parameters grows with the volume mapped, so the ultimate goal for the
cosmology community is to map our entire horizon volume, the region
from which light has had time to reach us during the 14 billion years
since our Big Bang.  Figure~\ref{fig:spheres} illustrates that whereas
the CMB mainly probes a thin shell from $z \sim 1000$, and current
large-scale structure probes (like galaxy clustering, gravitational
lensing, type Ia supernovae and the Lyman $\alpha$ forest) only map
small volume fractions nearby, neutral hydrogen tomography is able
to map most of our horizon volume.

\begin{table*}
\caption{Factors that affect the cosmological parameter measurement accuracy. \label{tab:methodology}}
\begin{ruledtabular}
\begin{tabular}{p{1.9cm}|p{2.8cm}|p{4.1cm}|p{4.1cm}|p{4.1cm}}
\multicolumn{2}{c|}{Assumptions}  &  Pessimistic & Middle & Optimistic \\ \hline
Power modeling & Ionization power spectrum modeling &  
  Marginalize over arbitrary $\Pmuzero$ and $\Pmutwo$ & 
  Marginalize over constants that parametrize $\sPxx(k)$ and $\sPxd(k)$ & 
  No ionization power spectrum, $\sPdd(k) \propto P_{\Delta T}(\bfk)$.  \\ \cline{2-5}

 &Non-linear cut-off scale $\kmax$ & $1\,\perMpc$ & $2\,\perMpc$  &$ 4\,\perMpc$ \\ \cline{2-5}

 & Non-Gaussianity of ionization signals & Doubles sample variance & \multicolumn{2}{c}{Negligible} \\  \hline

Cosmological & Reionization history & 
  \multicolumn{2}{c|}{ Gradual reionization over wide range of redshifts} & 
  Abrupt reionization at $z \lesssim 7$
   \\ \cline{2-5}
 &Redshift range  &7.3-8.2&  	$6.8 - 8.2$    &6.8 - 10\\ \cline{2-5}
 & Parameter space & 
Vanilla model plus optional parameters & 
       &
Vanilla model parameters\\ \hline   
   
Experimental & 
 Data & MWA, LOFAR, 21CMA & Intermediate case  &  SKA, FFTT \\ \cline{2-5}
 & Array configuration \footnotemark[1] & $\eta=0.15$ & $\eta=0.8$, $n=2$ & Giant core \\ \cline{2-5}
 & Collecting area \footnotemark[2]  &  $0.5\>\times$ design values & Design values & $2\>\times$ Design values \\ \cline{2-5}
 & Observation time \footnotemark[3]  & $1000\,{\rm hours}$  & $4000\,{\rm hours}$ & $16000\,{\rm hours}$ \\ \cline{2-5}
 & System temperature & $2\times T_{\rm sys}$ in \cite{bowman05cr}   &  $T_{\rm sys}$ given in \cite{bowman05cr}   & $0.5\times T_{\rm sys}$ in \cite{bowman05cr}   \\ \cline{2-5}
  \hline

Astrophysical & 
Residual foregrounds cut-off scale $\kmin$\footnotemark[4]  & $4\pi / yB$ & $2\pi / yB$ & $\pi / yB$ \\ 
\end{tabular}	
\end{ruledtabular}
\footnotetext[1]{For the FFTT, we consider only the case where all dipoles are in a giant core.}
\footnotetext[2]{See designed or assumed values of $A_e$ in Table \ref{tab:spec}.}
\footnotetext[3]{Assumes observation of two places in the sky.}
\footnotetext[4]{It is hard to predict the level of the residual foregrounds after the removal procedure. 
To quantify contributions from other factors, we take the approximation that there is no residual foregrounds at $k>\kmin$.  Here in the table, $yB$ is the comoving (l.o.s.) distance width of a single $z$-bin.}
\end{table*}

\begin{table*}
\tiny{ 
\caption{\label{tab:omp} The dependence of cosmological constraints on the full range of assumptions.  
We assume the fiducial values given in Section \ref{sec:cosmology}, and employ the Fisher matrix 
formalism to forecast the $\onesig$ accuracy of 21cm tomography measurements.   
Unless otherwise noted, errors are computed by marginalizing over all other parameters in the 
first ten columns (which we refer to as the ``vanilla'' parameters). 
In ``All OPT/MID/PESS'', we use the assumptions of the right, middle and left column of Table~\ref{tab:methodology}, respectively.
We assume that the total observing time is split between two sky regions, each for an amount in Table \ref{tab:methodology},  
using a giant/quasi-giant/small core array configuration where
100\%/80\%/15\%  of the antennae in the inner core are compactly laid at the array center while the rest 0\%/20\%/85\% of antennae 
fall off in density as $\rho\sim r^{-2}$ outside the compact core.  
}
\begin{ruledtabular}
\begin{tabular}{llccccccccccccc}
    &      & \multicolumn{10}{c}{\it Vanilla Alone} &  & &   \\ \cline{3-12}
		&  		    & $\Delta\Ol$ & 
 $\Delta\ln(\Omega_m h^2)$   & $\Delta\ln(\Omega_b h^2)$ & 
 $\Delta\ns$		       & $\Delta\ln\As$ & 
 $\Delta\tau$	               & $\Delta\xH(7.0)$ \footnotemark[1] & $\Delta\xH(7.5)$ & 
 $\Delta\xH(8.0)$            & $\Delta\xH(9.2)$ & $\Delta\Ok$ &
 $\Delta\mnu$ [\eV]	       & $\Delta\al$ \\ \hline
Planck    &          & 0.0070&0.0081&0.0059&0.0033&0.0088&0.0043&...&... &...&... & 0.025  & 0.23  & 0.0026 \\ \hline
          & All OPT  &0.0044&0.0052&0.0051&0.0018&0.0087&0.0042&0.0063&0.0063&0.0063&0.0063&0.0022&0.023&0.00073 \\
\:+LOFAR  & All MID  &0.0070&0.0081&0.0059&0.0032&0.0088&0.0043&0.18&0.26&0.23& ... &0.018&0.22&0.0026 \\
	  & All PESS &0.0070&0.0081&0.0059&0.0033&0.0088&0.0043& ... &51&49& ... &0.025&0.23&0.0026 \\ \hline
	  & All OPT  &0.0063&0.0074&0.0055&0.0024&0.0087&0.0043&0.0062&0.0062&0.0062&0.0062&0.0056&0.017&0.00054 \\
\:+MWA	  & All MID  &0.0061&0.0070&0.0056&0.0030&0.0087&0.0043&0.32&0.22&0.29 & ... &0.021&0.19&0.0026 \\
	  & All PESS &0.0070&0.0081&0.0059&0.0033&0.0088&0.0043& ... &29&30& ...&0.025&0.23&0.0026 \\ \hline
	  & All OPT  &0.00052&0.0018&0.0040&0.00039&0.0087&0.0042&0.0059&0.0059&0.0059&0.0059&0.0011&0.010&0.00027 \\
\:+SKA	  & All MID  &0.0036&0.0040&0.0044&0.0025&0.0087&0.0043&0.0094&0.014&0.011 & ... &0.0039&0.056&0.0022 \\
	  & All PESS &0.0070&0.0081&0.0059&0.0033&0.0088&0.0043& ...&1.1&1.0& ...&0.025&0.23&0.0026  \\ \hline
	  & All OPT  &0.00010&0.0010&0.0029&0.000088&0.0086&0.0042&0.0051&0.0051&0.0051&0.0051&0.00020&0.0018&0.000054 \\
\:+FFTT\footnotemark[2] &All MID & 0.00038&0.00034&0.00059&0.00033&0.0086&0.0042&0.0013&0.0022&0.0031&... &0.00023&0.0066&0.00017 \\
          & All PESS & 0.0070&0.0081&0.0059&0.0033&0.0088&0.0043&...&0.0043&0.0047& ...&0.025&0.11&0.0024 \\	
	  
\end{tabular} 
\end{ruledtabular}
\footnotetext[1]{$\xH(z)$ refers to the mean neutral fraction at redshift $z$.}
\footnotetext[2]{FFTT stands for Fast Fourier Transform Telescope, a future square kilometer array optimized for 21 cm tomography as described in \cite{FFTT}.  Dipoles in FFTT are all in a giant core, and this configuration does not vary. }
}
\end{table*}

Several recent studies have forecast the precision with which such 21
cm tomography can constrain cosmological parameters, both by mapping
diffuse hydrogen before and during the reionization epoch
\cite{McQuinn:2005hk,Bowman:2005hj,Santos:2006fp} and by mapping
neutral hydrogen in galactic halos after reionization
\cite{Wyithe:2007rq}.  These studies find that constraints based on
the cosmic microwave background measurements can be significantly
improved.  However, all of these papers make various assumptions, and
it is important to quantify to what extent their forecasts depend on
these assumptions.  This issue is timely because 21 cm experiments
(like LOFAR \cite{LOFAR}, 21CMA \cite{21CMA}, MWA \cite{MWA} and SKA
\cite{SKA}) are still largely in their planning, design or
construction phases.  These experiments will be described in detail 
in Section \ref{sec:experiments}.  
In order to maximize their scientific ``bang for
the buck'', it is therefore important to quantify how various design
tradeoffs affect their sensitivity to cosmological parameters.

The reason that neutral hydrogen allows mapping in three rather than
two dimensions is that the redshift of the $21$ cm line provides the
radial coordinate along the line-of-sight (l.o.s.).  This signal can
be observed from the so-called dark ages \cite{shapiro05,Lewis:2007kz} 
before any stars had formed,
through the epoch of reionization (EoR), and even to the current epoch
(where most of the neutral hydrogen is confined within galaxies).  We
focus in this study on the $21$ cm signal from $6 < z < 20$ -- the end
of the dark ages through the EoR.  This is the redshift
range at which the synchrotron foregrounds are smallest, and
consequently is the range most assessable for all planned $21$ cm
arrays. 

There are three position-dependent quantities that imprint signatures
on the 21 cm signal: the hydrogen density, the neutral fraction, and
the spin temperature. For cosmological parameter measurements, only
the first quantity is of interest, and the last two are nuisances.
(For some astronomical questions, the situation is reversed.)  The 21
cm spin-flip transition of neutral hydrogen can be observed in the
form of either an absorption line or an emission line against the CMB
blackbody spectrum, depending on whether the spin temperature is lower
or higher than the CMB temperature.

During the epoch of reionization, the spin temperature is likely coupled
to the gas temperature through Ly$\alpha$ photons via the
Wouthuysen-Field Effect \cite{wout,field58}, and the gas in the 
inter-galactic medium (IGM) has been
heated by X-ray photons to hundreds of Kelvin from the first stars
\cite{Pritchard:2007}.  If this is true, the 21cm signal will only
depend on the hydrogen density and the neutral fraction.  However,
astrophysical uncertainties prevent a precise prediction for exactly
when the gas is heated to well above the CMB temperature and is
coupled to the spin temperature.  In this paper, we follow
\cite{McQuinn:2005hk,Bowman:2005hj} and focus entirely on the regime
when the spin temperature is much larger than the CMB temperature
\cite{zald04,furlanetto04a,santos05}, such that the observed signal
depends only on fluctuations in density and/or the neutral fraction.
Specifically, we focus on the time interval from when this
approximation becomes valid (around the beginning of the reionization
\cite{zald04,furlanetto04a,santos05}) until most hydrogen has become
ionized, illustrated by the darkest region in
Figure~\ref{fig:spheres}.  Despite this simplification, the methods
that we apply to model the ionization fluctuations almost certainly can be
applied to model spin temperature fluctuations with minimal
additional free parameters.

In Table \ref{tab:methodology}, we list all the assumptions that affect the accuracy of cosmological parameter measurements, including ones about power
modeling, cosmology, experimental design, and astrophysical foregrounds. 
For each case, we provide three categories of assumptions:
one pessimistic (PESS), one middle-of-the-road (MID) 
and one optimistic (OPT). Since we wish to span the entire range of
uncertainties, we have made both the PESS and OPT models rather extreme.
The MID model is intended to be fairly realistic, but somewhat
on the conservative (pessimistic) side.

Before describing these assumptions in detail in the next section, it is important to note that taken together, they make a huge difference.
Table~\ref{tab:omp} illustrates this by showing the cosmological parameter constraints resulting from
using all the OPT assumptions, all the MID assumptions or all the PESS assumptions,  respectively.  
For example, combining CMB data from Planck and 21 cm data from FFTT,
the $\onesig$ uncertainty differs by a factor of 125 for $\Ok$ and by
a factor of 61 for $\mnu$ depending on assumptions.  It is therefore
important to sort out which of the assumptions contribute the most to
these big discrepancies, and which assumptions do not matter
much. This is a key goal of our paper.

The rest of this paper is organized as follows.  In Section
\ref{sec:assumptions}, we explain in detail the assumptions in the
same order as in Table \ref{tab:methodology}, and also present a new
method for modeling the ionization power spectra.  In Section
\ref{sec:results}, we quantify how the cosmological parameter
measurement accuracy depends on each assumption, and we derive simple
analytic approximations of these relations.  In Section
\ref{sec:conclusion}, we conclude with a discussion of the relative
importance of these assumptions, and implications for experimental
design.

\section{Forecasting Methods \& Assumptions}\label{sec:assumptions}

\subsection{Fundamentals of 21cm cosmology}
\subsubsection{Power spectrum of 21 cm radiation}\label{sec:review}

We review the basics of the 21 cm radiation temperature and power spectrum only briefly here, and refer the
interested reader to \cite{Furlanetto:2006jb} for a more comprehensive discussion of the relevant physics.
The difference between the observed 21 cm brightness temperature at the
redshifted frequency $\nu$ and the CMB temperature \Tcmb\ is \cite{field59a}
\beq{eqn:tb1}
T_b(\bfx) = \frac{3 c^3 h A_{10} n_H(\bfx) [\Ts(\bfx)-\Tcmb]}{32\pi k_B \nu_0^2 \Ts(\bfx) (1+z)^2 (dv_{\parallel}/dr)}\,,
\een
where \Ts\ is the spin temperature, $n_H$ is the number density of the neutral hydrogen gas, and
$A_{10}\approx 2.85\times 10^{-15}$s$^{-1}$ is the spontaneous decay rate of 21cm transition. The factor
$dv_{\parallel}/dr$ is the gradient of the physical velocity along the line of sight ($r$ is the
comoving distance), which is $H(z)/(1+z)$  on average (\ie for no peculiar velocity).  Here $H(z)$ is the
Hubble parameter at redshift $z$. The spatially averaged brightness temperature at redshift $z$ is (in units of $\milliK $)
\ben
\bar{T_b} \approx 23.88 \xH \left(\frac{\bar{\Ts} -\Tcmb}{\bar{\Ts}}\right) 
            \left(\frac{\Ob h^2}{0.02}\right) \left(\frac{0.15}{\Om h^2} \frac{1+z}{10} \right)^{1/2}\,,
\een
where $\xH$ is the mean neutral fraction, and $\bar{\Ts}$ is the averaged spin temperature.
If $\Ts \gg \Tcmb$ in the EoR, the 21cm emission should therefore be observed at the level of milli-Kelvins.  

To calculate the fluctuations, we rewrite \Eq{eqn:tb1} in terms of
$\delta$ (the hydrogen mass density fluctuation), $\delta_x$ (the
fluctuation in the ionized fraction), and the gradient of the peculiar
velocity $\partial v_r / \partial r$ along the line of sight, using
the fact that $dv_{\parallel}/dr = H(z)/(1+z) + \partial v_r /
\partial r$: 
\bena 
T_b(\bfx) &=& \tilde{T}_b \left[
1-\xI(1+\delta_x)\right] (1+\delta) \left(1-\frac{1}{Ha}\frac{\partial
v_r }{ \partial r }\right) \nonumber\\ 
& & \times \left(\frac{\Ts -\Tcmb}{\Ts}\right) \,. \label{eqn:tb2}
\eena Here $\xI\equiv 1-\xH$ is the mean ionized fraction, and we have
defined $\tilde{T}_b\equiv \bar{T}_b/\xH\times [\bar{\Ts}/(\bar{\Ts}
-\Tcmb)]$.  We write $\delta_v \equiv (Ha)^{-1} \partial v_r /
\partial r$. In Fourier space, it is straightforward to show that, as
long as $\delta\ll 1$ so that linear perturbation theory is valid,
$\delta_v (\bfk) = -\mu^2 \delta$, where $\mu=\hat{\bfk}\cdot\hat{{\bf
n}}$ is the cosine of the angle between the Fourier vector $\bfk$ and
the line of sight.  In this paper, we restrict our attention to the
linear regime.  We will also throughout this paper assume $\Ts \gg
\Tcmb$ during the EoR, making the last factor in \Eq{eqn:tb2} unity
for the reasons detailed in Section \ref{sec:intro}.

In Fourier space, the power spectrum $\PDT(\bfk)$ of the 21cm fluctuations is defined by $\langle \Delta T_b^{\,*}(\bfk)  \Delta T_b(\bfk ') \rangle
\equiv  (2\pi)^3\delta^3 (\bfk-\bfk ') \PDT(\bfk)$, where $\Delta T_b$ is the deviation from the mean brightness
temperature.  It is straightforward to show from \Eq{eqn:tb2} that, to leading order,
\bena
 \PDT(\bfk) & = & \tilde{T}_b^2 \left\{ [\xH^2 \Pdd -2\xH \Pxd + \Pxx] \right. \nonumber \\
 & & \left. + 2\mu^2 [\xH^2 \Pdd - \xH \Pxd ] + \mu^4 \xH^2 \Pdd \right\}\,. \label{eqn:21cmpower}
\eena
Here $\Pxx = \xI^2 P_{\delta_x \delta_x}$ and $\Pxd = \xI P_{\delta_x \delta}$ are the ionization power spectrum and the density-ionization power spectrum respectively.   
For convenience, we
define $\sPdd(k)\equiv \tilde{T}_b^2 \xH^2 \Pdd(k)$, $\sPxd(k) \equiv \tilde{T}_b^2 \xH \Pxd(k)$ and $\sPxx(k) \equiv \tilde{T}_b^2
\Pxx(k)$, so the total 21 cm power spectrum can be written as
three terms with different angular dependence:
\beq{eqn:fourmoments}
\PDT(\bfk) = \Pmuzero(k) + \Pmutwo(k) \mu^2 + \Pmufour(k) \mu^4,
\een
where 
\bena
\Pmuzero&=&\sPdd -2 \sPxd + \sPxx,\label{eqn:pmu0}\\ 
\Pmutwo&=&2(\sPdd - \sPxd),\label{eqn:pmu2}\\ 
\Pmufour&=&\sPdd.\label{eqn:pmu4}
\eena
Since $\Pmufour$ involves only the matter
power spectrum that depends only on cosmology, 
Barkana and Loeb \cite{barkana04a,barkana05} argued that in principle, one can separate cosmology from astrophysical
``contaminants'' such as $\sPxx$ and $\sPxd$ whose physics is hitherto far from known. 
We will quantify the accuracy of this conservative approach (which corresponds to our PESS scenario for ionization
power spectrum modeling below) in Section \ref{sec:results}.

\subsubsection{From ${\bf u}$ to ${\bf k}$}

The power spectrum $\PDT(\bfk)$ and the comoving 
 vector ${\bf k}$ (the Fourier dual of the comoving position vector ${\bf r}$) are not directly measured by 21cm experiments.  
An experiment cannot directly determine which position vector ${\bf r}$ a signal is coming from, but instead which 
vector ${\bf \Theta} \equiv \theta_x \hat{e}_x +\theta_y \hat{e}_y + \Delta f \hat{e}_z$ it is coming from, where 
$(\theta_x,\theta_y)$ give the angular location on the sky plane, and $\Delta f$ is the frequency difference from the central redshift of a $z$-bin.
For simplicity, we assume that the sky volume observed is small enough that we can linearize 
the relation between ${\bf \Theta}$ and ${\bf r}$.
Specifically, we assume that the sky patch observed is much less than a radian across, 
so that we can approximate the sky as flat 
\footnote{The FFTT is designed for all-sky mapping (i.e. the field of view is of order $2\pi$).  However, 
since the angular scales from which we get essentially all our cosmological information are much smaller than a radian 
(with most information being on arcminute scales), the flat-sky approximation is accurate as long as
the data is analyzed separately in many small patches and the constraints are subsequently combined.}, 
and that separations in frequency near the mean redshift $z_*$ are approximately proportional to separations in comoving distance.
In these approximations, if there are no peculiar velocities, 
\bena
{\bf\Theta}_\perp	&=&\frac{{\bf r}_\perp}{d_A(z_*)} \,\label{eqn:thetaperp},\\ 
\Delta f		 &=&\frac{\Delta r_\parallel}{y(z_*)}.\label{eqn:thetaz}
\eena
Here ``$\perp$'' denotes the vector component perpendicular to the line of sight, 
{\it i.e.}, in the $(x,y)$-plane, and
$d_A$ is the comoving angular diameter distance given by \cite{KolbTurnerBook}
\ben\label{dAEq}
d_A(z)={c\over H_0}|\Omega_k|^{-1/2}S\left[|\Omega_k|^{1/2}\int_0^z \frac{dz'}{E(z')}\right],
\eeq
where 
\beq{EdefEq}
E(z)\equiv \frac{H(z)}{H_0} = \sqrt{\Om(1+z)^3 + \Ok(1+z)^2+\Ol},
\eeq
is the relative cosmic expansion rate and
the function $S(x)$ equals $\sin (x)$ if $\Ok<0$, $x$ if $\Ok=0$, and $\sinh x$ if $\Ok>0$. 
The conversion factor between comoving distances intervals and frequency intervals is 
\ben\label{yDefEq} 
y(z) = \frac{\lambda_{21}(1+z)^2}{H_0 E(z)},
\een
where $\lambda_{21}\approx 21$ cm is the rest-frame wavelength of the 21 cm line.  

We write the Fourier dual of ${\bf \Theta}$ as $\bfu \equiv u_x \hat{e}_x + u_y \hat{e}_y  + \upar \hat{e}_z$ ($\upar$ has units of time).  
The relation between $\bfu$ and $\bfk$ is therefore
\bena
\uperp&=&d_A {\bf k}_\perp\,,\label{eqn:uperp}\\ 
\upar &=&y\,\kpar\,.\label{eqn:uz}
\eena
In $\bfu$-space, the power spectrum $\PDT(\bfu)$ of 21cm signals is defined by 
$\langle \Delta \tilde{T}^{*}_b(\bfu) \Delta \tilde{T}_b(\bfu ')\rangle=(2\pi)^3 \delta^{(3)}(\bfu-\bfu')  \PDT(\bfu)  $,
and is therefore related to $\PDT(\bfk)$ by 
\beq{PuPkEq}
\PDT(\bfu)= {1\over d_A^2 y} \PDT(\bfk)\,.
\een

Note that cosmological parameters affect $\PDT(\bfu)$ in two ways: 
they both change $\PDT(\bfk)$ and alter
the geometric projection from ${\bf k}$-space to ${\bf u}$-space.
If $d_A$ and $y$ changed while $\PDT(\bfk)$ remained fixed, the observable power spectrum
$\PDT(\bfu)$ would be dilated in both the $\uperp$ and $\upar$ directions and rescaled in amplitude,
while retaining its shape.
Since both $d_A$ and $y$ depend on the three parameters  
$(\Ok,\Ol,h)$, and the Hubble parameter is in turn given by the parameters in Table~\ref{tab:omp}
via the identity $h=\sqrt{\Omega_m h^2/(1-\Ol-\Ok)}$, we see that these geometric effects provide information only
about our parameters $(\Ok,\Ol,\Omega_m h^2)$. 
Baryon acoustic oscillations in the power spectrum provide a powerful ``standard ruler'',
and the equations above show that if one generalizes to the dark energy to make $\Ol$ an arbitrary function of $z$, then 
the cosmic expansion history $H(z)$ can be measured separately at each 
redshift bin, as explored in \cite{Wyithe:2007rq,Mao:2007ti,Chang:2007xk}.
21 cm tomography information on our other cosmological
parameters ($\ns$, $\As$, $\Omega_b h^2$, $m_\nu$, $\alpha$, etc.) 
thus comes only from their direct effect on $\PDT(\bfk)$. 
Also note that $(\Ok,\Ol)$ affect $\PDT(\bfk)$ only by modulating the rate of linear perturbation growth,
so they alter only the amplitude and not the shape of $\PDT(\bfk)$.

If we were to use \Eq{PuPkEq} to infer $\PDT(\bfk)$ from the measured power spectrum $\PDT(\bfu)$
while assuming incorrect cosmological parameter values, then this geometric scaling
would cause the inferred $\PDT(\bfk)$ to be distorted by the so-called Alcock-Paczy\'nski (AP) effect
\cite{nusser04,barkanaAP} and not take the simple form of
Eqns.(\ref{eqn:fourmoments})-(\ref{eqn:pmu4}).
To avoid this complication, we therefore perform our Fisher matrix analysis directly in terms of 
$\PDT(\bfu)$, since this quantity is directly measurable without any cosmological assumptions.

The above transformations between ${\bf u}$-space and ${\bf r}$-space are valid when there are no peculiar velocities.
The radial peculiar velocities $v_r$ that are present in the real world induce the familiar redshift space distortions
that were discussed in Section \ref{sec:review}, causing $\mu^2$ and $\mu^4$ power spectrum anisotropies that were described there.

\subsection{Assumptions about \sPxx\ and \sPxd }
\label{sec:ion-model}

During the EoR, ionized bubbles (HII regions) in the IGM grow and
eventually merge with one another.  Consequently, \sPxx(k) and
\sPxd(k) contribute significantly to the total 21cm power spectrum.
The study of the forms of these two ionization power spectra has made
rapid progress recently, particularly through the semi-analytical
calculations \cite{furlanetto04a,zald04,mcquinn05,Zahn:2005fn} and radiative
transfer simulations \cite{McQuinn:2007dy,Zahn:2006sg}.  However, these models
depend on theoretically presumed parameters whose values cannot
currently be calculated from first principles.  From the experimental
point of view, it is therefore important to develop data analysis
methods that depend only on the most generic features of the
ionization power spectra. In this paper, we consider three methods ---
our OPT, MID and PESS models --- that model $\sPxx$ and $\sPxd$ as
follows: \beq{eqn:allP0} {\rm (OPT)} \qquad\qquad \left\{
\begin{array}{lcl}
\sPxx(k) & = & 0 \\ 
\sPxd(k) & = & 0
\end{array}
\right.
\een

\[ {\rm (MID)} \qquad\qquad\qquad\qquad\qquad\qquad \]
\ben\label{eqn:parn_pxxd}
\left\{ 
\begin{array}{lcl}
\sPxx(k) & = & \bsqxx \left[ 1+\alxx(k\,\Rxx) + \,(k\,\Rxx)^2\right]^{- {\gaxx \over 2}} \sPdd^{\rm (fid)}\\ 
\sPxd(k) & = & \bsqxd \,\exp{\left[-\alxd (k\,\Rxd)-(k\,\Rxd)^2\right]} 		 \sPdd^{\rm (fid)}
\end{array}
\right.
\een

\ben
{\rm (PESS)} \qquad\qquad
\left\{ 
\begin{array}{lcl}
\sPxx(k) & = & {\rm arbitrary} \\ 
\sPxd(k) & = & {\rm arbitrary}
\end{array}
\right.
\een
In the next three subsections, we explain these models in turn.

\subsubsection{OPT model}

It is likely that before reionization started (while $\xH=1$ and $\sPxx=\sPxd=0$), 
hydrogen gas had already been sufficiently heated that $\Ts \gg \Tcmb$. 
In this regime, \Eq{eqn:allP0} holds. This OPT scenario
is clearly the simplest model, since the total 21cm power spectrum is simply proportional to $\sPdd$:
$\PDT(\bfk)=\sPdd(k) (1+\mu^2)^2$.  To forecast the $\onesig$ error, we use the Fisher matrix
formalism \cite{tegmark97b}.  
Repeating the derivation in \cite{galfisher},
the Fisher matrix for cosmological parameters $\lambda_a$ ($a=1,\ldots,N_p$) is 
\beq{FisherIntegralEq} 
{\bf F}_{ab} = \frac{1}{2}\int 
\left(\frac{\partial\ln\PDT^{\rm tot}(\bfu)}{\partial\lambda_a}\right)  
\left(\frac{\partial\ln\PDT^{\rm tot}(\bfu)}{\partial\lambda_b}\right) 
V_{\Theta} \frac{d^3 u}{(2\pi)^3},
\eeq
where the integral is taken over the observed part of ${\bf u}$-space, 
and $\PDT^{\rm tot}(\bf u)$ denotes the combined power spectrum from cosmological 
signal and all forms of noise.  Here $V_{\Theta} = \Omega \times B$ is the volume of
 the ${\bf \Theta}$-space where $\Omega$ is the solid angle within the field of view (f.o.v.) and $B$ is the frequency size of a $z$-bin.  
The Fisher matrix determines the parameter errors as $\Delta\lambda_a = \sqrt{({\bf F}^{-1})_{aa}}$. 

For computational convenience, we subdivide $u$-space into pixels so small that the 
power spectrum is roughly constant in each one,
obtaining 
\beq{eqn:FM4cp} 
{\bf F}_{ab}\approx\sum_{\rm pixels} \frac{1}{[\delta \PDT(\bfu)]^2}\left(\frac{\partial \PDT(\bfu)}{\partial \lambda_a}\right)  \left(\frac{\partial \PDT(\bfu)}{\partial \lambda_b}\right),
\een
where the power spectrum measurement error in a pixel at $\bfu$ is 
\beq{OPT_P_err_eq}
\delta\PDT(\bfu)=\frac{\PDT^{\rm tot}(\bfu)}{N_c^{1/2}} =  \frac{\PDT(\bfu)+P_N(u_\perp)}{N_c^{1/2}}.
\eeq
Here $P_N(u_\perp)$ is the noise power spectrum and will be discussed in detail in Section \ref{sec:noise};
note that it is independent of $u_\parallel$ and depends only on $u_\perp$ through the
baseline distribution of the antenna array.
\beq{NcDefEq}
N_c=2\pi k^2\sin\theta \Delta k \Delta\theta \times {\rm Vol}/(2\pi)^3
\eeq
is the number of independent cells in an annulus summing over
the azimuthal angle.  We have the factor $\sqrt{1/N_c}$ in $\delta\PDT$ instead of the normal $\sqrt{2/N_c}$ because we only
sum over half the sphere.

\begin{table}
\footnotesize{
\caption{\label{tab:ri-fid} Fiducial values of ionization parameters adopted for Figure \ref{fig:ri}.  $\Rxx$ and $\Rxd$ are in units of $\Mpc$, while other parameters are unitless. }
\begin{ruledtabular}
\begin{tabular}{lcccccccc}

$z\quad\quad$     & \xH  & \bsqxx   & \Rxx     &  \alxx   & 
\gaxx    & \bsqxd  & \Rxd     &  \alxd   \\ \hline

9.2  & 0.9 & 0.208 & 1.24 & -1.63 & 0.38 & 0.45 & 0.56  & -0.4 \\
8.0  & 0.7 & 2.12  & 1.63 & -0.1  & 1.35 & 1.47 & 0.62  & 0.46 \\
7.5  & 0.5 & 9.9   & 1.3  & 1.6   & 2.3  & 3.1  & 0.58  & 2.   \\
7.0  & 0.3 & 77.   & 3.0  & 4.5   & 2.05 & 8.2  & 0.143 & 28.  \\

\end{tabular}
\end{ruledtabular}
}
\end{table}

\begin{figure*}[ht]
\centering
\begin{displaymath}
\begin{array}{cccc} 
  \includegraphics[width=0.25\textwidth]{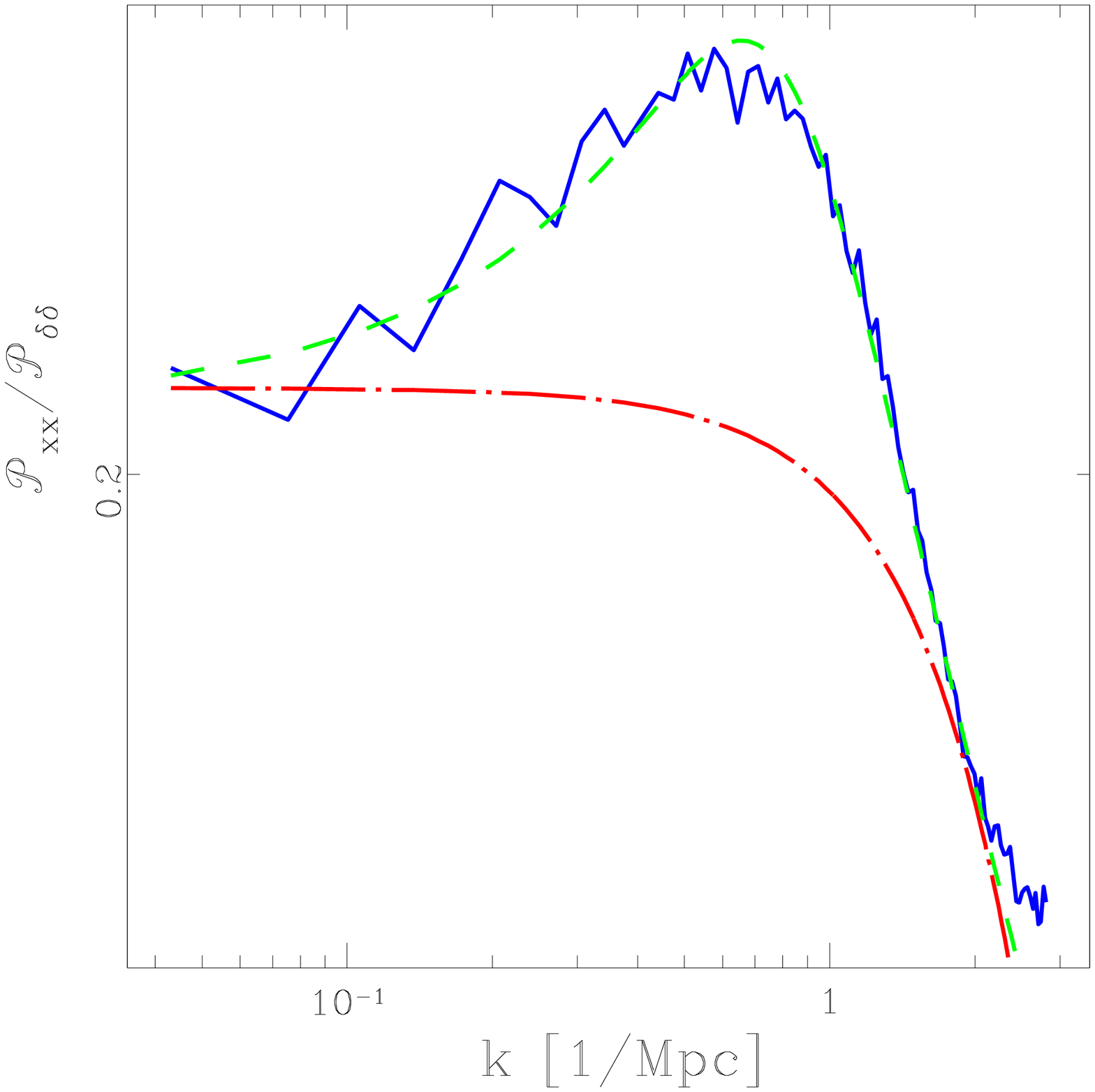} & 
  \includegraphics[width=0.25\textwidth]{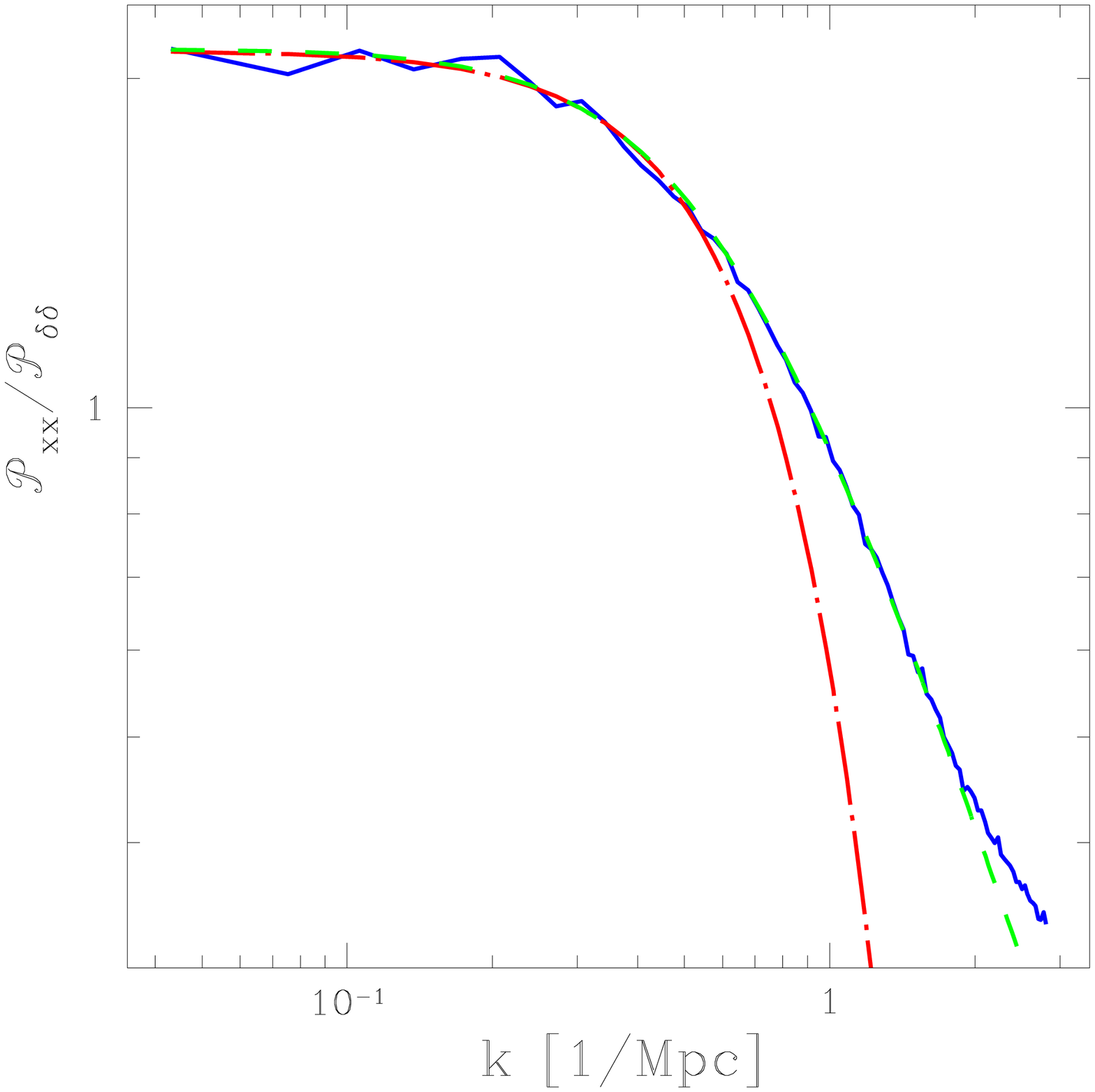} & 
  \includegraphics[width=0.25\textwidth]{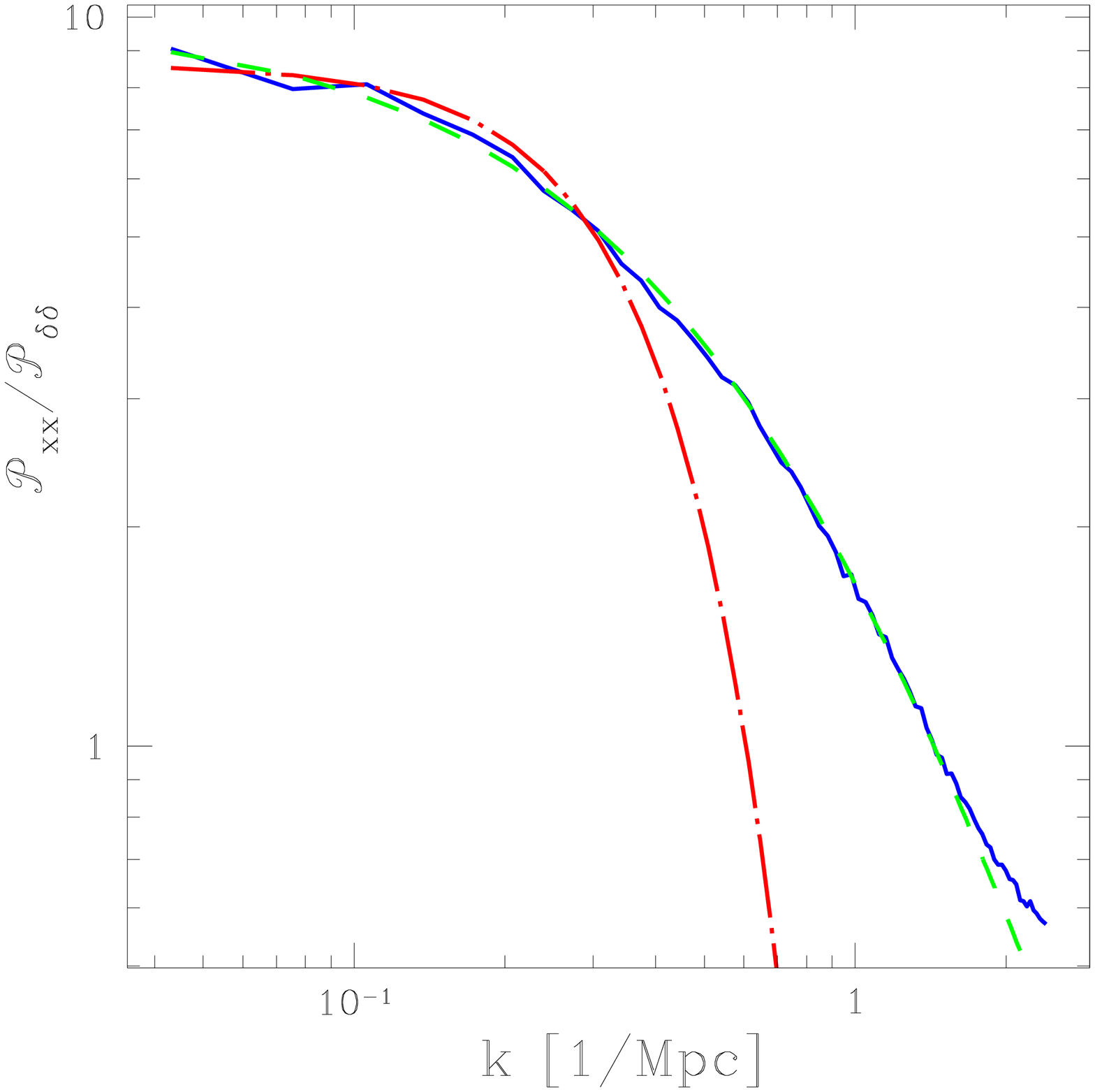} & 
  \includegraphics[width=0.25\textwidth]{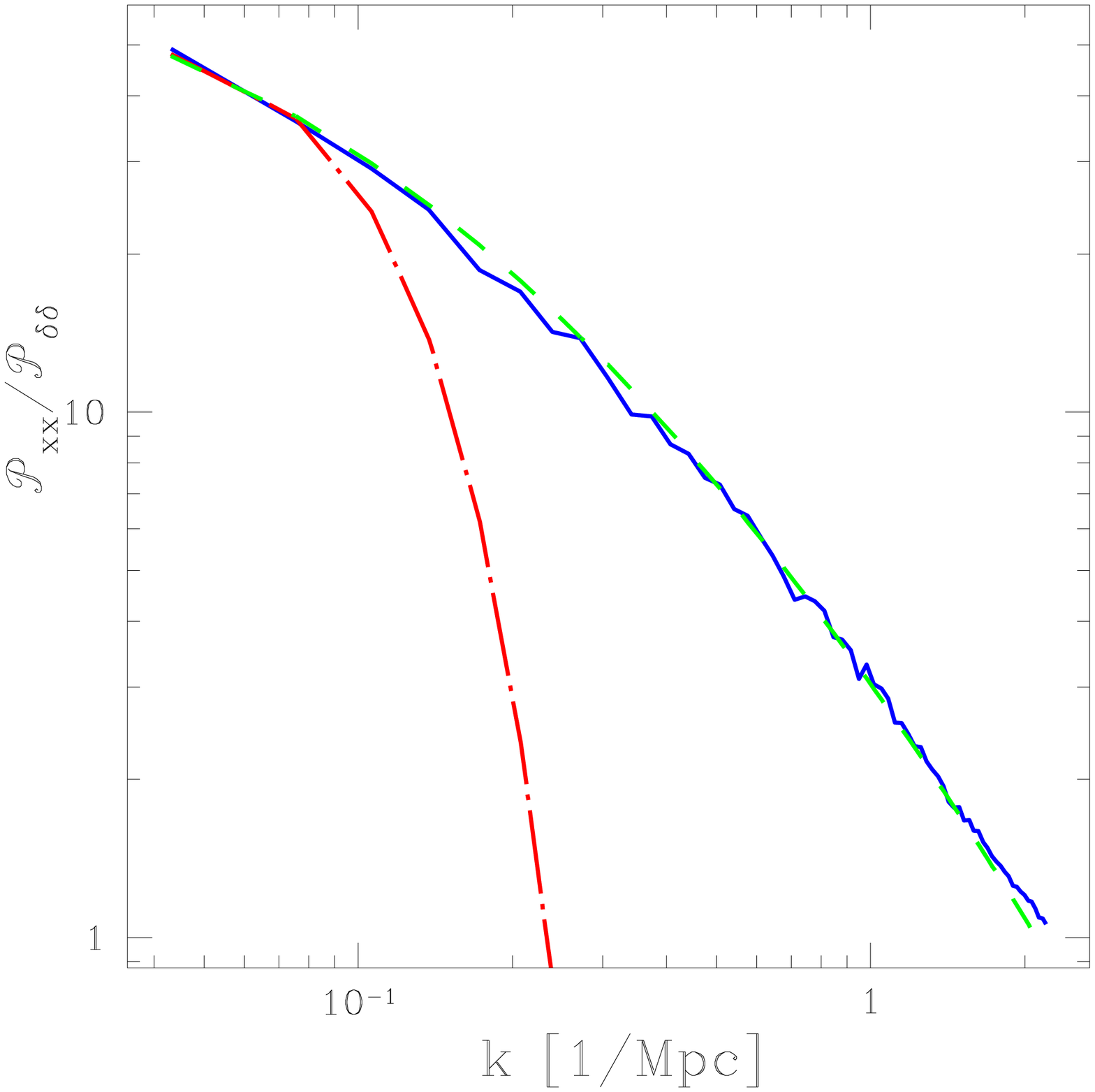} \\
  \includegraphics[width=0.25\textwidth]{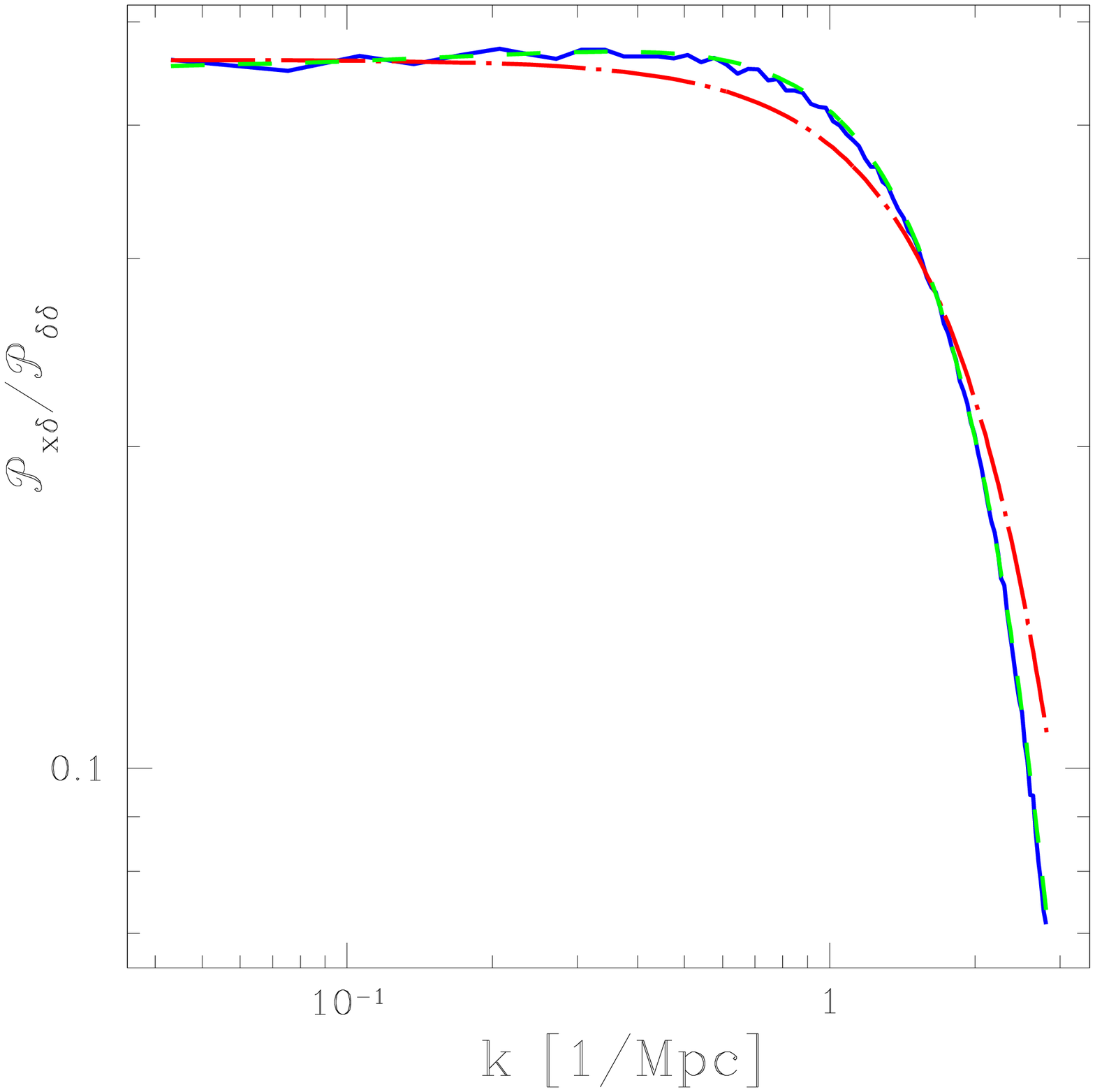} &  
  \includegraphics[width=0.25\textwidth]{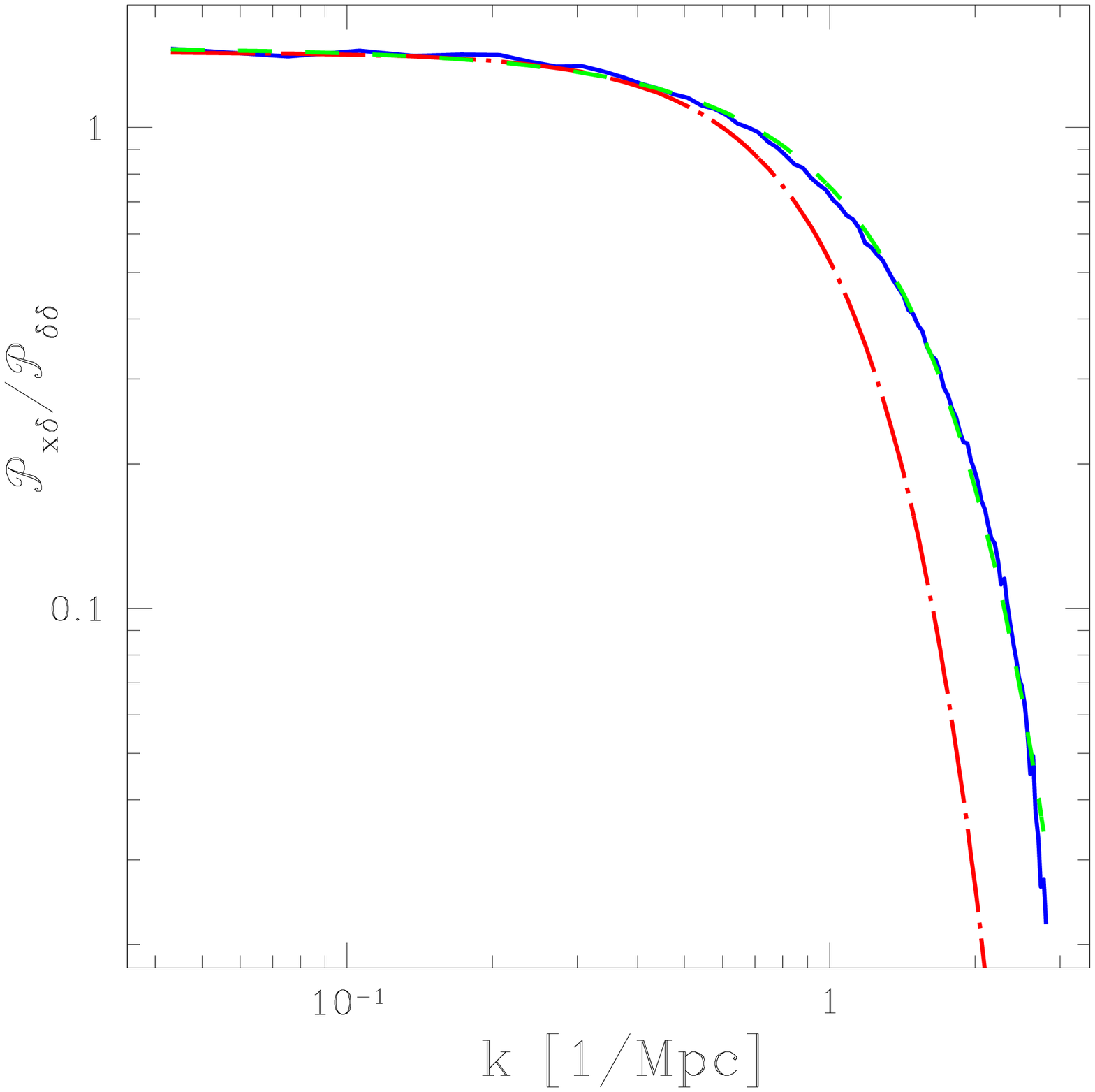} &  
  \includegraphics[width=0.25\textwidth]{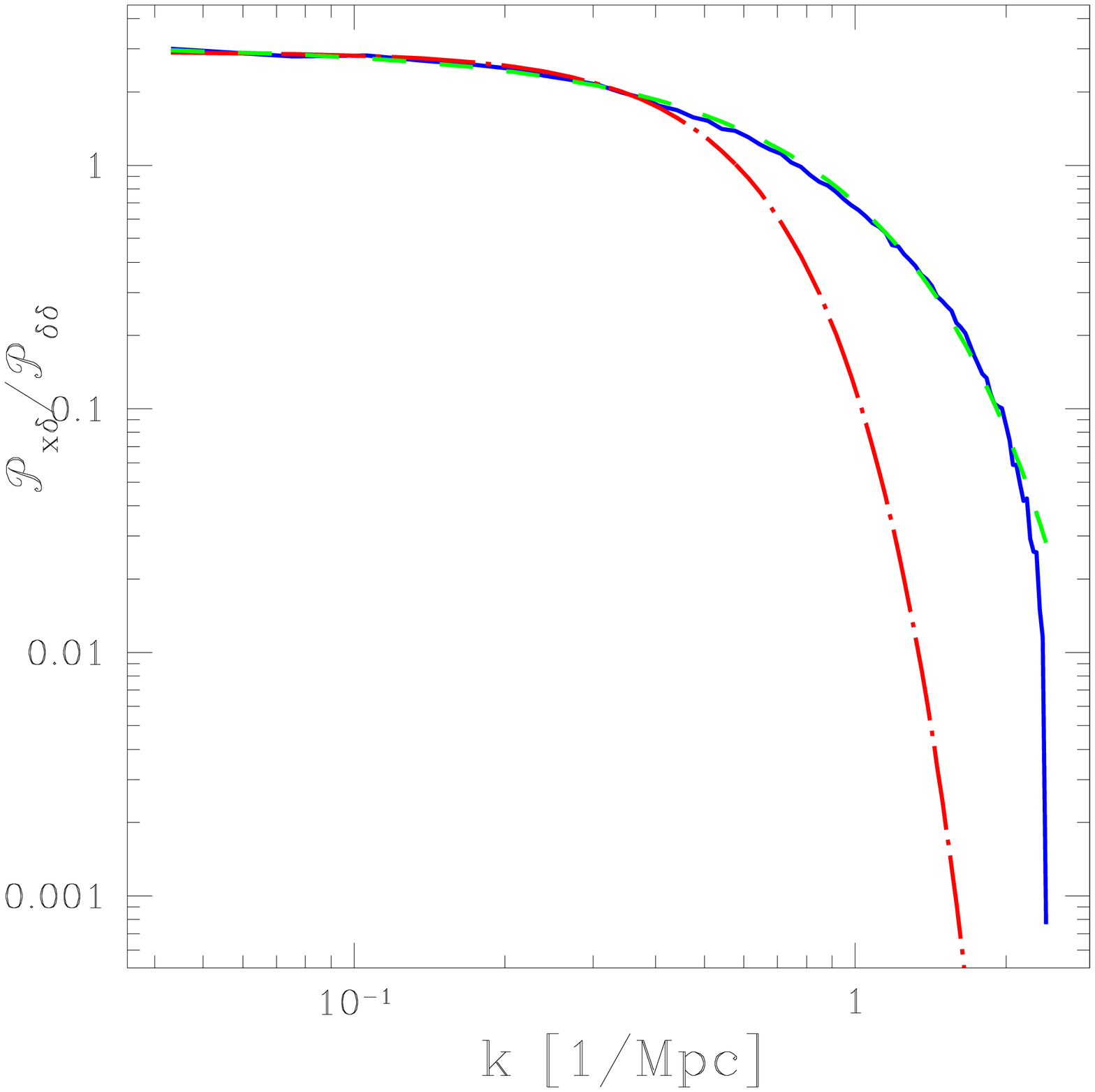} &  
  \includegraphics[width=0.25\textwidth]{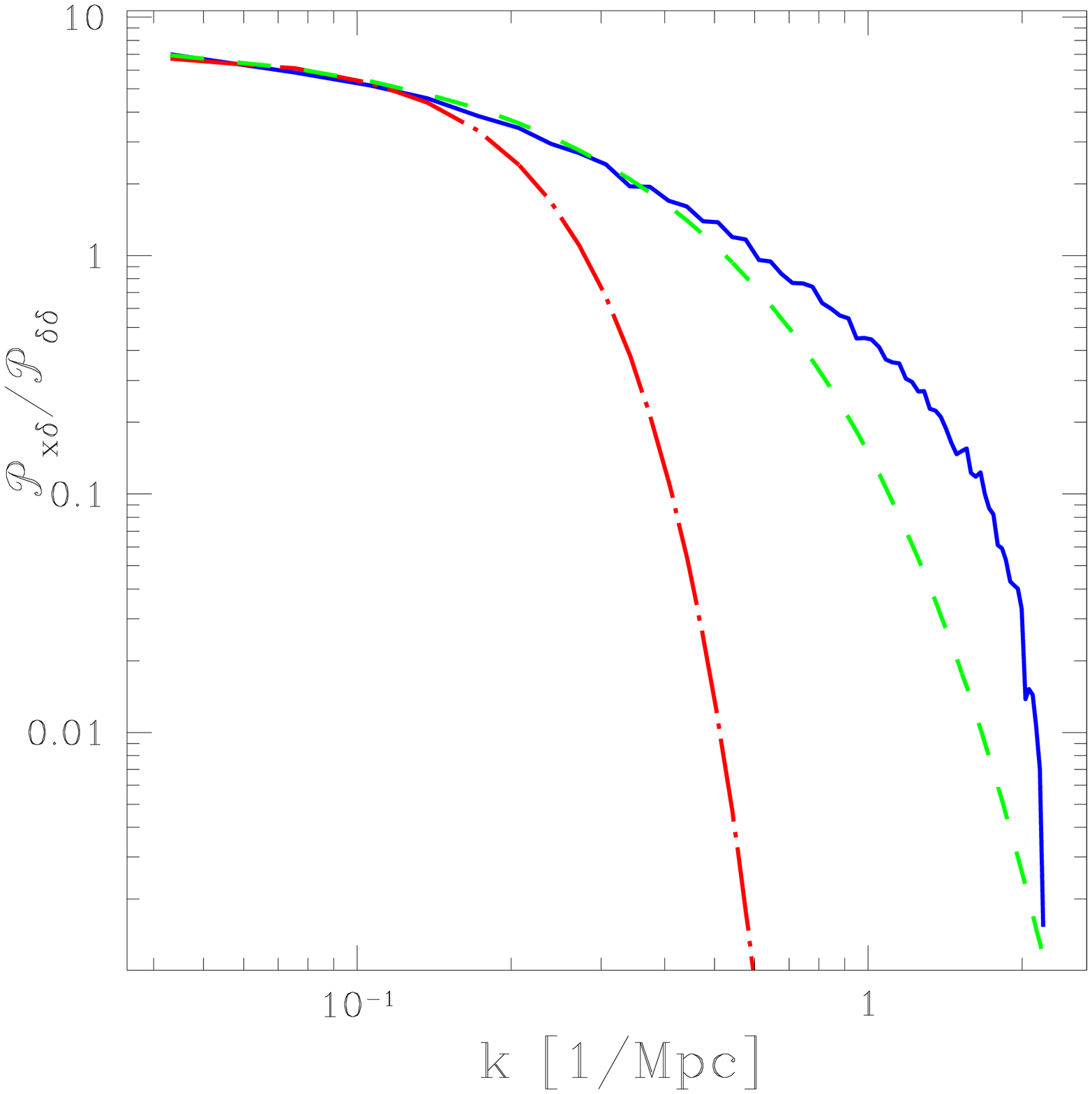} 
\end{array}
\end{displaymath}
\caption{ Fits to the ionization power spectra at several redshifts.
Solid (blue) lines are the results of the radiative transfer
simulation in Model I of the McQuinn \etal paper
\cite{McQuinn:2007dy}.  Dashed (green) lines are fitting curves of our
parametrization. Dot-dashed (red) lines are best fits using the
parametrization suggested by Santos and Cooray \cite{Santos:2006fp} .
Top panels: $ \sPxx/\sPdd=\Pxx/(\xH^{2} \Pdd)$.  Bottom panels: $
\sPxd/\sPdd=\Pxd/(\xH \Pdd)$.  From left to right:
$z=9.2,\,8.0,\,7.5,\,7.0$ ($\xI=0.10,\, 0.30,\, 0.50,\, 0.70$
respectively).}
\label{fig:ri}
\end{figure*}

\subsubsection{MID model}\label{sec:asmp-mid}

After reionization starts, both ionization power spectra $\sPxx$ and
$\sPxd$ make significant contribution to the total 21cm power
spectrum.  We explore two different analysis methods --- our MID and
PESS models --- for separating the cosmological signal from these
astrophysical contaminants (\ie $\sPxx$ and $\sPxd$).

Our MID model assumes that both ionization power spectra $\sPxx(k)$
and $\sPxd(k)$ are smooth functions of $k$ which can be parametrized
by a small number of nuisance parameters $\beta_1,\ldots,\beta_{n_{\rm
ion}}$ related to reionization physics.  Combining these ionization
parameters with our cosmological ones $\lambda_a$ into a larger
parameter set $p_\alpha$ ($\alpha=1,\ldots,N_p+n_{\rm ion}$), we can
jointly constrain them by measuring $\PDT(\bfu)$.

In Appendix \ref{chi2} we will describe a $\chi^2$ goodness-of-fit
test for quantifying whether this parametrization is valid.  The
Fisher matrix for measuring $p_\alpha$ is simply \beq{MIDfisherEq}
F_{\alpha\beta} = \sum_{\rm pixels}
\frac{1}{[\delta\PDT(\bfu)]^2}\frac{\partial \PDT(\bfu)}{\partial
p_\alpha}\frac{\partial \PDT(\bfu)}{\partial p_\beta} \,.  \eeq This
Fisher matrix $F_{\alpha\beta}$ is not block diagonal, {\it i.e.},
there are correlations between the cosmological and ionization
parameters, reflecting the fact that both affect $\sPxx(k)$ and
$\sPxd(k)$.  The inversion of the Fisher matrix therefore leads to the
degradation of the constraints of cosmological parameters.  However,
the total 21cm power spectrum is usually smaller in magnitude in the
MID model than in the OPT model (see \Eq{eqn:21cmpower}), giving less
sample variance.  This means that as long as noise in a 21cm
experiment dominates over sample variance, the MID model will give
weaker constraints than the OPT model, because of the degeneracies.
For future experiments with very low noise, however, it is possible to
have the opposite situation, if the reduction in sample variance
dominates over the increase in degeneracy.  This does of course not
mean that the MID model is more optimistic than the OPT model, merely
that the OPT model is assuming an unrealistic power spectrum.

Having set up the general formalism, we now propose a parametrization specified 
by \Eq{eqn:parn_pxxd}, with fiducial values
of ionization parameters given in Table \ref{tab:ri-fid}.  This
parametrization was designed to match the results of the radiative
transfer simulations in Model I of \cite{McQuinn:2007dy}, and
Figure~\ref{fig:ri} shows that the fit is rather good in the range
$k=0.1-2 \; \Mpc^{-1}$ to which the 21cm experiments we consider are most
sensitive.

The radiative transfer simulations implemented in
\cite{McQuinn:2007dy} are post processed on top of a $1024^3$ N-body
simulation in a box of size $186\,\Mpc$. Three models for the
reionization history are considered in \cite{McQuinn:2007dy}:
\begin{enumerate}
\item In Model I, all dark matter halos above $m_{\rm cool}$
(corresponding to the minimum mass galaxy in which the gas can cool by
atomic transitions and form stars, \eg $m_{\rm cool}\approx
10^8\,M_{\odot}$ at $z=8$) contribute ionizing photons at a rate that
is proportional to their mass.
\item In Model II, the ionizing luminosity of the sources scales as halo mass to
the $5/3$ power, \ie more massive halos dominate the production of ionizing photons than in Model I.
\item In Model III, which has the same source parametrization as in
Model I except for doubled luminosity, minihalos with $m>10^5 M_{\odot}$ absorb incident ionizing photons out to their virial radius unless they are
photo-evaporated (but do not contribute ionizing photons).
\end{enumerate}

It appears to be a generic feature in the simulation results that the ratios of functions at large $k$ fall off like a power law for
$\sPxx(k)/\sPdd(k)$, and exponentially for $\sPxd(k)/\sPdd(k)$.  At small $k$, $\sPxx(k)/\sPdd(k)$ can either increase or decrease
approximately linearly as $k$ increases, while $\sPxd(k)/\sPdd(k)$ is asymptotically constant.  Our parametrization in
\Eq{eqn:parn_pxxd} captures these features: at large $k$, $\sPxx(k)/\sPdd(k) \propto k^{-\gaxx}$ and $\sPxd(k)/\sPdd(k) \propto \exp{(-(k\,\Rxd)^2)}$; at small $k$, $\sPxx(k)/\sPdd(k) \propto \left(1- (\gaxx \, \alxx\Rxx/2)\,k\right)$, and $\sPxd(k)/\sPdd(k)
\propto \left(1- \alxd\,\Rxd\,k \right)$ (both $\alxx$ and $\alxd$ can be either positive or negative).
Figure~\ref{fig:ri} also shows that for $\Pxx(k)$ and also for $\Pxd(k)$ at large $k$, 
our parametrization further improves over the 
parametrization $P(k)/\Pdd = b^2 e^{-(k\,R)^2}$
suggested by Santos and Cooray \cite{Santos:2006fp}, which works well for $\Pxd(k)$ at small $k$.

To be conservative, we discard cosmological information from $\sPxd(k)$ and $\sPxx(k)$ in our Fisher matrix analysis by using the fiducial power spectrum $\sPdd(k)^{\rm (fid)}$ rather than the actual one $\sPdd(k)$ in \Eq{eqn:parn_pxxd}.
This means that the derivatives of $\sPxd(k)$ and $\sPxx(k)$ with respect to the cosmological parameters vanish in \Eq{MIDfisherEq}.
It is likely that we can do better in the future: once  
the relation between the ionization power spectra and the matter power spectrum can be reliably calculated either analytically or numerically, 
the ionization power spectra can contribute to further constraining cosmology.    

In addition to the fit of Model I shown in Figure \ref{fig:ri}, we also fit our model (with different fiducial values from those listed in Table \ref{tab:ri-fid}) to the simulations using Model II and III in \cite{McQuinn:2007dy}, and find that the parametrization is flexible enough to provide good fits to all three simulations,  
suggesting that the parametrization in
\Eq{eqn:parn_pxxd} may be generically valid and independent of models.  
Note, however, that at low redshifts ($\xI \gtrsim 0.7$), our parametrization of $\sPxd/\sPdd$  does not work well at large $k$, in that the
simulation falls off less  rapidly than exponentially.  This may be because when HII regions dominate the IGM, the ionized bubbles  overlap in complicated
patterns and correlate extremely non-linearly at small scales.  This partial incompatibility indicates that our parametrization (\ie Eq.\ref{eqn:parn_pxxd}) is only accurate for small
$\xI$, \ie before non-linear ionization patterns come into play.

In the remainder of this paper, we will adopt the values in Table \ref{tab:ri-fid} as  fiducial values of the ionization parameters.

\subsubsection{PESS model}

By parametrizing the ionization power spectra with a small number of
constants, the MID model rests on our understanding of the physics of
reionization.  From the point of view of a maximally cautious
experimentalist, however, constraints on cosmological parameters
should not depend on how well one models reionization. In this spirit,
Barkana and Loeb \cite{barkana04a,barkana05} proposed what we adopt as our
``PESS'' model for separating the physics $\sPdd(k)$ from the
``gastrophysics'' $\sPxx(k)$ and $\sPxd(k)$. Instead of assuming a
specific parametrization, the PESS model
makes no {\it a priori} assumptions about the ionization power spectra.
In each $k$-bin that contains more than three pixels in ${\bf u}$-space, one can in principle separate $\Pmufour(k)=\sPdd(k)$ from the other two moments.  
The PESS model essentially only constrains cosmology from the $\Pmufour$ term and therefore loses all information in $\Pmuzero$ and $\Pmutwo$.  
We now set up the Fisher matrix formalism for the PESS model that takes advantage of the anisotropy in $\PDT(\bfk)$ arising from the velocity field effect.  Numerical
evaluations will be performed in Section \ref{sec:result-power}.  

The observable in 21cm tomography is the brightness temperature $T_b(\bfx)$. In Fourier space, the covariance matrix between two pixels $\bfk_i$ and $\bfk_j$ is
${\bf C}_{ij}=\delta_{ij}[\PDT(\bfk_i)+P_N(k_{\perp})]$, assuming that the measurements in two different pixels are uncorrelated\footnote{We ignore here a 
$\delta$-function centered at the origin since 21cm experiments will not measure any $k=0$ modes.}.     
The total 21cm power spectrum is $\PDT(\bfk) = \Pmuzero(k) + \Pmutwo(k)
\mu^2 + \Pmufour(k) \mu^4$.  For convenience, we use the shorthand notation $P_A$, where $P_1\equiv\Pmuzero$, 
$P_2\equiv\Pmutwo$ and $P_3\equiv \Pmufour$ and define
the $a_A=0,2,4$ for $A=1,2,3$, respectively.  Thus the power spectrum can be rewritten as $\PDT=\sum_{A=1}^{3} P_A \mu^{a_A}$.  
Treating $P_A(k)$ at each $k$-bin as
parameters, the derivatives of the covariance matrix are simply $\partial {\bf C}_{ij}/ \partial P_A(k) = \delta_{ij} \mu^{a_A}$, 
where $|\bfk_i|$ resides in the shell of radius
$k$ with width $\Delta k$. 
Since the different $k$-bins all decouple, the Fisher matrix 
for measuring the moments $P_A(k)$ is simply a separate $3\times 3$-matrix for each $k$-bin:
\bena
F_{AA'} (k) & = & \frac{1}{2}\hbox{tr}\left[{\bf C}^{-1}\frac{\partial {\bf C}}{\partial P_A(k)}{\bf C}^{-1}\frac{\partial {\bf C}}{\partial P_{A'}(k)}\right] \nonumber\\
& = &\sum_{\textrm{upper half-shell}} \frac{\mu^{a_A + a_{A'}}}{[\delta \PDT(\bfk)]^2}\,, \label{eqn:Fpess}
\eena
where $\delta \PDT(\bfk) = N_c^{-1/2}\left[ \PDT(\bfk) + P_N(k_{\perp})\right]$.  
Here $P_N(k_{\perp})$ is related to $P_N(u_\perp)$ by \Eq{PuPkEq}.  
Again the sum is over the upper half of the spherical shell $k<|\bfk|<k+\Delta k$.  
The $\onesig$ error of $P_3=\Pmufour $ is $\delta P_3(k) = \sqrt{F^{-1}_{\phantom{-1}33}(k)}$. 
Once $\sPdd=\Pmufour$ is separated from other moments, $\sPdd$ can be used to constrain cosmological parameters $\lambda_a$ with the Fisher matrix as given in \Eq{eqn:FM4cp}.

We have hitherto discussed the anisotropy in $\PDT(\bfk)$ that arises from the velocity field effect.
However, the AP-effect may further contribute to the anisotropy in that it
creates a $\mu^6$-dependence and modifies the $\mu^4$ term \cite{nusser04,barkanaAP}.  
The AP-effect can be distinguished from the velocity field effect since the $\Pmusix$ term is unique to the AP-effect. 
Thus, one can constrain cosmological parameters from $\Pmufour$ and $\Pmusix$ \cite{McQuinn:2005hk}, 
involving the inversion of a $4\times 4$ matrix which loses even more information and therefore further weakens constraints.  
Therefore, the PESS Fisher matrix that we have derived without taking AP-effect into account can be viewed as an upper bound on how 
well the PESS approach can do in terms of cosmological parameter constraints.
However, this maximally conservative $4\times 4$ matrix approach may be inappropriately pessimistic, since the AP-induced clustering anisotropy is typically very small within the observationally allowed
cosmological parameter range, whereas the velocity-induced anisotropies can be of order unity.

\subsection{Assumptions about Linearity}

To avoid fitting to modes where $\delta_k$ is non-linear and physical
modeling is less reliable, we impose a sharp cut-off at $\kmax$ and
exclude all information for $k>\kmax$. We take $\kmax = 2\,\perMpc$
for our MID model, and investigate the $\kmax$-dependence of
cosmological parameter constraints in Section \ref{sec:kmax}.  

\subsection{Assumptions about non-Gaussianity}
\label{sec:non-gauss}

Non-Gaussianity of ionization signals generically becomes important at
high $\xI$. Using cosmic reionization simulations with a large volume 
and high resolution, Lidz \etal \cite{Lidz:2006vj} and Santos \etal
\cite{Santos:2007dn} found non-negligible (a factor of 1.5)
differences in the full power spectrum at high $\xI$ ($\xI \gtrsim
0.35)$).  To get a rough sense of the impact of non-Gaussianity on
cosmological parameter constraints, we simply model it as increasing
the sample variance by a factor $\xi$.  We thus write the total power
spectrum as \beq{xiDefEq}
\delta\PDT(\bfu)=N_c^{-1/2}\,[\,\xi\PDT(\bfu)+P_N(u_\perp)]\,, \eeq
where $\xi$ is the factor by which the the sample variance is
increased.  The parameter $\xi$ should take the value $\xi \approx 1$
(Gaussian) at epochs with low $\xI$ and $1 < \xi \lesssim 2$
(non-Gaussian) at high $\xI$.

\subsection{Assumptions about reionization history and redshift range}
\label{sec:asmp-history}

21cm tomography can probe a wide range of redshifts, as illustrated in
Figure~\ref{fig:spheres}.  However, one clearly cannot simply measure
a single power spectrum for the entire volume, as the clustering
evolves with cosmic time: The matter power spectrum changes gradually
due to the linear growth of perturbations \cite{tegmark05}.  More importantly, the
ionization power spectra vary dramatically with redshift through the
epoch of reionization.  We incorporate these complications by
performing our analysis separately in a series of redshift slices,
each chosen to be narrow enough that the matter and ionization power
spectra can be approximated as constant in redshift within each slice.  This
dictates that for a given assumed reionization history, thinner
redshift slices must be used around redshifts where $\xH$ varies
dramatically.

In this paper, we will consider two rather opposite toy models in Section \ref{sec:results}:
\begin{itemize}
\item OPT: A sharp reionization that begins and
finishes at one redshift (say $z\lesssim 7$).
\item MID/PESS: A gradual reionization that spans a range of redshifts,
assuming the ionization parameter values that fit Model I simulation of  the McQuinn \etal paper \cite{McQuinn:2007dy}
\end{itemize}
For the latter scenario, the ionization fraction $\xH$ is not a linear function of redshift. For example, in 
in the McQuinn \etal \cite{McQuinn:2007dy} simulation, $\xH=$0.9, 0.7, 0.5 and 0.3 correspond to redshifts 
$z=9.2$, 8.0, 7.5 and 7.0, respectively.
For our different scenarios, we therefore adopt the redshift ranges $6.8<z<10$ that are divided into four redshift slices centered at the above redshifts (OPT),
$6.8<z<8.2$ split into three bins centered at $z$=7.0, 7.5 and 8.0 (MID), 
$7.3<z<8.2$ split into two slices centered at $z=7.5$ and 8.0. 

\subsection{Assumptions about cosmological parameter space}\label{sec:cosmology}

Since the impact of the choice of cosmological parameter space and related degeneracies has been extensively studied in the literature,
we will perform only a basic analysis of this here.
We work within the context of standard inflationary cosmology with adiabatic perturbations, and parametrize cosmological models in terms of 12
parameters (see, \eg, Table 2 in \cite{lrg_astro-ph_0608632} for explicit definitions) whose fiducial values are assumed as follows: 
$\Ok = 0$ (spatial curvature), $\Ol = 0.7$ (dark energy density), $\Ob = 0.046$ (baryon density),
$h = 0.7$ (Hubble parameter $H_0 \equiv 100 h \,{\rm km}\,{\rm s}^{-1}\,{\rm Mpc}^{-1}$), $\tau = 0.1$ (reionization optical depth), $\On=0.0175$
(massive neutrino density), $\ns=0.95$ (scalar spectral index), $A_s = 0.83$ (scalar fluctuation amplitude), $r=0$ (tensor-to-scalar ratio), $\al=0$
(running of spectral index), $n_t=0$ (tensor spectral index) and $w=-1$ (dark energy equation of state).  
We will frequently use the term ``vanilla'' to refer to the minimal model space parametrized by $(\Ol,\om,\ob,\ns,\As,\tau)$ combined with $\xH(z)$ and
ionization parameters at all observed $z$-bins, setting $\Ok,\on,r,\al,\nt$, and $w$ fixed at their fiducial values.  

\begin{table}
\footnotesize{
\caption{Specifications for 21cm interferometers\label{tab:spec}}
\begin{ruledtabular}
\begin{tabular}{c|p{1cm}|p{1.5cm}|p{1.5cm}|p{2cm}}

Experiment & $N_{\rm ant}$ & Min. baseline (m) & f.o.v. (${\rm deg}^2$) & $A_e$ (m$^2$) at z=6/8/12\footnote{We 
assume that the effective collecting area is proportional to $\lambda^2$ 
such that the
sensitivity ($A_e/T_{\rm sys}$ in ${\rm m}^2 {\rm K}^{-1}$) meets the design specification.}  \\ \hline
MWA   & 500 & 4 & $\pi\,16^2$ & 9/14/18 \\ 
SKA   & 7000 & 10 & $\pi\,8.6^2$ & 30/50/104 \\
LOFAR & 77 & 100 & $2\times\pi\,2.4^2$ & 397/656/1369 \\
FFTT  & $10^6$ & 1 & $2\pi$ & 1/1/1 \\
\end{tabular}
\end{ruledtabular}

}
\end{table}

\subsection{Assumptions about Data} \label{sec:experiments}

The MWA, LOFAR, SKA and FFTT instruments are still in their planning/design/development stages. 
In this paper, we adopt the key design parameters from \cite{bowman05cr}
for MWA, \cite{Schilizzi07} and www.skatelescope.org for SKA, www.lofar.org for LOFAR, and \cite{FFTT} for FFTT unless explicitly stated.  

\subsubsection{Interferometers} 

We assume that MWA will have 500  
correlated $4{\rm m} \times 4{\rm m}$ antenna tiles, each with 16
dipoles.  Each individual tile will have an effective collecting area
of $14 \,{\rm m}^2$ at $z=8$ and $18\, {\rm m}^2$ at $z\gtrsim 12$.
LOFAR will have 77 large (diameter $\sim 100 \,{\rm m}\,) $ stations,
each with thousands of dipole antennae such that it has the collecting
area nearly 50 times larger than each antenna tile of MWA.  Each
station can simultaneously image $N$ regions in the sky.  We set $N=2$
in this paper but this number may be larger for the real array.  The
design of SKA has not been finalized.  We assume the ``smaller
antennae'' version of SKA, in which SKA will have 7000 small antennae,
much like MWA, but each panel with much larger collecting area.  FFTT
stands for Fast Fourier Transform Telescope, a future square kilometer
array optimized for 21 cm tomography as described in \cite{FFTT}.
Unlike the other interferometers we consider, which add in phase the
dipoles in each panel or station, FFTT correlates all of its dipoles,
resulting in more information.  We evaluate the case where FFTT contains
a million $1{\rm m} \times 1{\rm m}$ dipole antennae in a contiguous
core subtending a square kilometer, providing a field-of-view of
$2\pi$ steradians.

For all interferometers, we assume that the collecting area $A_e \propto \lambda^2$, like a simple dipole, except that $A_e$ is saturated at $z\sim 12$ in MWA
since the wavelength $\lambda=21(1+z)\,{\rm cm}$ exceeds the physical radius of an MWA antenna panel.  The summary of the detailed specifications adopted in this
paper is listed in Table \ref{tab:spec}. 

\subsubsection{Configuration}
\label{configuration}

The planned configurations of the above-mentioned interferometers are
quite varied. However, all involve some combination of the following
elements, which we will explore in our calculations:
\begin{enumerate}
\item A {\it nucleus} of radius $R_0$ within which the area coverage fraction is close to 100\%.
\item A {\it core} extending from radius $R_0$ our to $R_{\rm in}$ where there coverage density drops like some power law $r^{-n}$.
\item An {\it annulus} extending from $R_{\rm in}$ to $R_{\rm out}$ where the coverage density is low but rather uniform.
\end{enumerate}
In its currently planned design, the MWA will have a small 
nucleus, while the core density falls off as $r^{-2}$
until a sharp cutoff at $R_{\rm in}$.  
For LOFAR we assume 
32 stations in the core, 
and another 32 stations in an outer annulus out to radius $R_{\rm out}\sim
6\,{\rm km}$.  For SKA we assume 20\% in the core, and 30\% in the
annulus out to radius $R_{\rm out}\sim 5\,{\rm km}$. We ignore the
measurements from any dilute distribution of antenna panels outside
$R_{\rm out}$. For LOFAR and SKA, we assume a uniform distribution of
antennae in the annulus, but with an inner core profile like that of
the MWA, i.e., a nucleus of radius $R_0=285/189\,{\rm m}$ (LOFAR/SKA) 
and an $r^{-2}$
fall-off outside this compact core.  We assume an azimuthally
symmetric distribution of baselines in all arrays.

For an array with $N_{\rm in}$ antennae within $R_{\rm in}$, we can define a
quantity
\ben
R_0^{\rm max} \equiv \sqrt{\frac{N_{\rm in}}{\rho_0 \pi}}\,,
\een
where $\rho_0$ is the area density of the nucleus.
$R_0^{\rm max}$ is the maximal radius of the nucleus, corresponding to the case where there it 
contains all the $N_{\rm in}$ antennae and there is no core.

It is also convenient to parametrize the distribution of these $N_{\rm in}$ antennae within $R_{\rm in}$ by two numbers: 
the fraction $\eta$ that are in the nucleus, and the fall-off index $n$ of the core.  
It is straightforward to show that $R_0$ and $R_{\rm in}$ are related to $\eta$ and $n$ by 
\ben
R_0 = \sqrt{\eta} R_0^{\rm max}\,,
\een
\beq{eqn:Rin-R0}
R_{\rm in} = R_0 \left( \frac{2-n(1-\eta)}{2\eta} \right)^{\frac{1}{2-n}}
\een
if $n\ne 2$. The analytic relation for $n=2$ is $R_{\rm in} = R_0 \exp{[(1-\eta)/(2\eta)]}$, which can be well approximated in numerical calculation by by taking $n=2+\epsilon$ in \Eq{eqn:Rin-R0} with $\epsilon\sim 10^{-10}$.  

\begin{figure}[ht]
\centering
\includegraphics[width=0.5\textwidth]{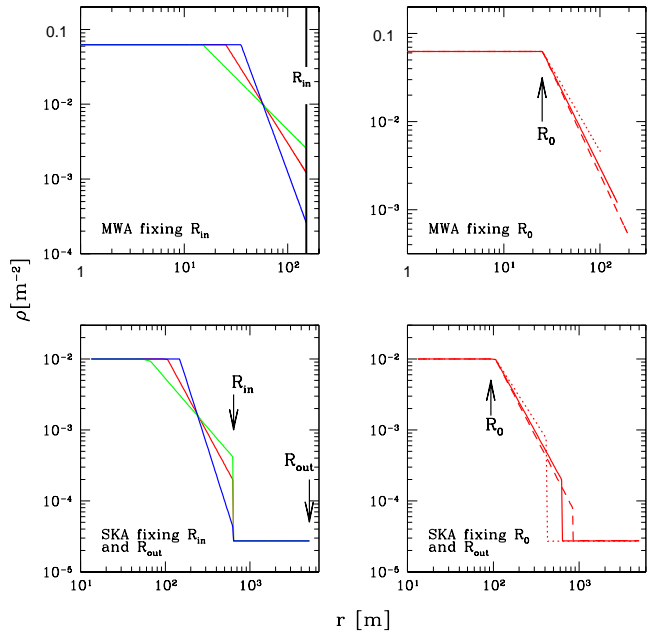} 
\caption{Examples of array configuration changes.  For MWA (upper panels), antennae are uniformly distributed inside the nucleus radius $R_0$, and the density $\rho$ falls off like a power law for $R_0<r<R_{in}$ where $R_{in}$ is the core radius.  For SKA (lower panels) and similarly for LOFAR, there is in addition a uniform yet dilute distribution of antennae in the annulus $R_{in}< r< R_{out}$, where $R_{out}$ is the outer annulus radius.  When $R_0$ is decreased ($R_0=0.7/0.5/0.3\times R_0^{\rm max}$) with $R_{in} = 3.0\times R_0^{\rm max}$ fixed (left panels), the density in the core falls off slower (blue/red/green curves).  When $R_{in}$ is decreased ($R_{in}=4.0/3.0/2.0\times R_0^{\rm max}$) with $R_0=0.5\times R_0^{\rm max}$ fixed (right panels), the density in the core also falls off less steep (dashed/solid/dotted curves).
}
\label{fig:layout}
\end{figure}

In Section \ref{sec:oc-results}, we will scan almost all possible
design configurations and find the optimal one for constraining
cosmology.  
There are two independent ways to vary array configurations, 
as illustrated by Figure \ref{fig:layout}: by varying $R_0$ with $R_{in}$ fixed, and by 
varying $R_{in}$ with $R_0$ fixed.  Contributions from antennae in 
the annulus are negligible compared to the core, so varying $R_{out}$ is 
not interesting.  

In other parts of Section \ref{sec:results}, we will
assume the intermediate configuration $\eta=0.8$ and $n=2$ 
(except for FFTT which is purely in a giant core) 
with the planned number of antennae in the core and annulus.  
Note that this configuration is optimized from the currently planned design.

\subsubsection{Detector noise} \label{sec:noise}  

21cm radio interferometers measure visibility $\bfV$.  The visibility
for a pair of antennae is defined as \cite{Morales:2004} \ben
\bfV(u_x,u_y,\Delta f) = \int dxdy \Delta T_b(x,y,\Delta f) e^{-i(u_x
x+u_y y)}\,, \een where $(u_x,u_y)$ are the number of wavelengths
between the antennae.  The hydrogen 3D map is the Fourier transform in
the frequency direction $\tilde{I}(\bfu)\equiv \int d\Delta f
\bfV(u_x,u_y,\Delta f) \exp{(-i\Delta fu_\parallel)}$ where $\bfu =
u_x\hat{e}_x+u_y\hat{e}_y+u_\parallel\hat{e}_z$.  The detector noise
covariance matrix for an interferometer is
\cite{Morales:2005,McQuinn:2005hk} \ben C^N(\bfu_i,\bfu_j) = \left(
\frac{\lambda^2 B T_{\rm sys}}{A_e}\right)^2 \frac{\delta_{ij}}{B
t_{\bfu_i}} \,, \een where $B$ is the frequency bin size, $T_{\rm
sys}$ is system temperature, and $t_{\bfu} \approx (A_e
t_0/\lambda^2)n(u_\perp)$ is the observation time for the visibility
at $|\bfu_\perp| = d_A|\bfk|\sin\theta$.  Here $t_0$ is the total
observation time, and $n$ is the number of baselines in an observing
cell.

The covariance matrix of the 21cm signal $\tilde{I}(\bfu)$ is related to the power spectrum $\PDT(\bfk)$ by \cite{McQuinn:2005hk}
\bena
C^{SV}(\bfu_i,\bfu_j) & \equiv & \langle \tilde{I}^{*}(\bfu_i) \tilde{I}(\bfu_j) \rangle\nonumber\\
 & = & \PDT(\bfu_i) \frac{\lambda^2 B}{A_e} \delta_{ij}\,.
\eena
Therefore, the noise in the power spectrum is 
\beq{eqn:PNoise}
P^{N}(u_\perp)=\left( \frac{\lambda^2 T_{\rm sys}}{A_e}\right)^2 \frac{1}{t_0 n(u_\perp)} \,.
\een

For all interferometers, the system temperature is dominated by sky temperature $T_{\rm sky} \approx 60(\lambda/1\,{\rm m})^{2.55}\,{\rm K}$ due to synchrotron radiation
in reasonably clean parts of the sky.  Following \cite{bowman05cr}, we set $T_{\rm sys} = 440\,K$ at $z=8$ and $T_{\rm sys} = 690\,K$ at $z=10$.

\subsection{Assumptions about Residual Foregrounds}\label{sec:foregrounds}

There have been a number of papers discussing foreground removal for 21 cm tomography (\eg  \cite{Wang:2005zj,DiMatteo:2004dt,Oh:2003jy,DiMatteo:2001gg} and references therein),  
and much work remains to be done on this important subject, as the 
the amplitudes of residual foregrounds depend strongly depends on cleaning techniques and assumptions, 
and can have potentially dominate the cosmological signal.
The work of Wang \etal \cite{Wang:2005zj} and McQuinn \etal \cite{McQuinn:2005hk} suggested that after fitting out a low-order polynomial from the frequency dependence in each pixel,
the residual foregrounds were negligible for $k> 2\pi/yB$ where $yB$ is the comoving width of a $z$-bin.  
To obtain a crude indication of the impact of residual foregrounds, there therefore we adopt the rough approximation that all data below some cutoff value $\kmin$ is destroyed by foregrounds while
the remainder has negligible contamination. We choose $\kmin=(1/2/4)\times \pi/yB$ for the OPT/MID/PESS scenarios, and also explore wider ranges below.

\section{Results and discussion}\label{sec:results}

In this section, we numerically evaluate how the accuracy of
cosmological parameter constraints depend on the various assumptions
listed above.  Where possible, we attempt to provide intuition for
these dependences with simple analytical approximations.  In most
cases, we explore the dependence on one assumption at a time by
evaluating the PESS, MID and OPT scenario for this assumption while
keeping all other assumptions fixed to the baseline MID scenario.

\subsection{Varying ionization power spectrum modeling and reionization histories} \label{sec:result-power}

\begin{table*}
\footnotesize{
\caption{\label{tab:power} How cosmological constraints depend on the ionization power spectrum modeling and reionization history.  We assume observations of 4000 hours on two places in the sky in the range of $z=6.8 - 8.2$ that is divided into three $z$-bins centered at $z=7.0$, $7.5$ and $8.0$ respectively, $\kmax = 2\perMpc$, $\kmin = 2\pi/yB$ and a quasi-giant core configuration (except for FFTT that is a giant core).  
$\onesig$ errors of ionization parameters in the MID model, marginalized over other vanilla parameters, are listed separately in \Tab{tab:mid2}.}
\begin{ruledtabular}

\begin{tabular}{llcccccccccccc}
    &      & \multicolumn{9}{c}{\it Vanilla Alone} &  & &   \\ \cline{3-11}
		&   Model    & $\Delta\Ol$ & 
 $\Delta\ln(\Omega_m h^2)$   & $\Delta\ln(\Omega_b h^2)$ & 
 $\Delta\ns$		       & $\Delta\ln\As$ & 
 $\Delta\tau$	               & $\Delta\xH(7.0)$ \footnotemark[1] & $\Delta\xH(7.5)$ & 
 $\Delta\xH(8.0)$            &  $\Delta\Ok$ &
 $\Delta\mnu$ [\eV]	       & $\Delta\al$ \\ \hline

LOFAR & OPT  &0.025&0.27&0.44&0.063&0.89&...&...&...&...&0.14&0.87&0.027   \\
      & MID  &0.13&0.083&0.15&0.36&0.80&...&...&...&...&0.35&12&0.17   \\ \cline{2-14}

MWA   & OPT &0.046&0.11&0.19&0.022&0.37&...&...&...&...&0.056&0.38&0.013  \\
      & MID &0.22&0.017&0.029&0.097&0.76&...&...&...&...&0.13&9.6&0.074  \\ \cline{2-14}
      
SKA   & OPT  &0.0038&0.044&0.083&0.0079&0.16&...&...&...&...&0.023&0.12&0.0040  \\
      & MID  &0.014&0.0049&0.0081&0.012&0.037&...&...&...&...&0.043&0.36&0.0060  \\ \cline{2-14}

      & OPT  &0.00015&0.0032&0.0083&0.00040&0.015&...&...&...&...&0.00098&0.011&0.00034 \\
FFTT  & MID  &0.00041&0.00038&0.00062&0.00036&0.0013&...&...&...&...&0.0037&0.0078&0.00017 \\
      & PESS &1.1&0.017&0.037&0.010&0.19&...&...&...&...&... &0.20&0.0058 \\	  
\hline

Planck &     & 0.0070 & 0.0081 & 0.0059 & 0.0033 & 0.0088 & 0.0043 & \nodata & \nodata & \nodata & 0.025  & 0.23  & 0.0026 \\ \cline{2-14}
         & OPT  &0.0066&0.0077&0.0058&0.0031&0.0088&0.0043&0.0077&0.0084&0.0093&0.0051&0.060&0.0022 \\
\:+LOFAR & MID  &0.0070&0.0081&0.0059&0.0032&0.0088&0.0043&0.18&0.26&0.23&0.018&0.22&0.0026 \\
         & PESS &0.0070&0.0081&0.0059&0.0033&0.0088&0.0043&0.54&0.31&0.24&0.025&0.23&0.0026   \\  \cline{2-14}

         & OPT  &0.0067&0.0079&0.0057&0.0031&0.0088&0.0043&0.0065&0.0067&0.0069&0.0079&0.027&0.0014  \\
\:+MWA   & MID  &0.0061&0.0070&0.0056&0.0030&0.0087&0.0043&0.32&0.22&0.29&0.021&0.19&0.0026  \\
         & PESS &0.0070&0.0081&0.0059&0.0033&0.0088&0.0043&3.8&0.87&0.53&0.025&0.23&0.0026   \\  \cline{2-14}
      
         & OPT  &0.0031&0.0038&0.0046&0.0013&0.0087&0.0042&0.0060&0.0060&0.0060&0.0017&0.017&0.00064   \\
\:+SKA   & MID  &0.0036&0.0040&0.0044&0.0025&0.0087&0.0043&0.0094&0.014&0.011&0.0039&0.056&0.0022 \\
         & PESS &0.0070&0.0081&0.0059&0.0033&0.0088&0.0043&0.061&0.024&0.012&0.025&0.21&0.0026   \\  \cline{2-14}

         & OPT  &0.00015&0.0015&0.0036&0.00021&0.0087&0.0042&0.0056&0.0056&0.0056&0.00032&0.0031&0.000094 \\
\:+FFTT  & MID  &0.00038&0.00034&0.00059&0.00033&0.0086&0.0042&0.0013&0.0022&0.0031&0.00023&0.0066&0.00017 \\
         & PESS &0.0055&0.0064&0.0051&0.0030&0.0087&0.0043&0.0024&0.0029&0.0040&0.025&0.020&0.0010 \\
     

\end{tabular}
\end{ruledtabular}
\footnotetext[1]{$\xH(z)$ denotes the mean neutral fraction at the central redshift $z$.  
$\xH(z)$'s and $A_s$ are completely degenerate from the 21cm measurement alone.  For this reason, the errors shown for $\ln\As$ from 21cm data alone is really not marginalized over $\xH(z)$'s.}
}
\end{table*}

\begin{table}
\footnotesize{
\caption{\label{tab:mid2} $\onesig$ marginalized errors for the ionization parameters in the MID model.  Assumptions are made the same as in \Tab{tab:power}.  $\Rxx$ and $\Rxd$ are in units of $\Mpc$ and other parameters are unitless. } 
\begin{ruledtabular}
\begin{tabular}{clccccccc}
$z$ &   &  $\Delta\bsqxx$	 &  $\Delta\Rxx$ &  $\Delta\alxx$ &  $\Delta\gaxx$   &  $\Delta\bsqxd$	 &  $\Delta\Rxd$ &  $\Delta\alxd$ \\ \hline
        & Values& 77.   & 3.0  & 4.5   & 2.05 & 8.2  & 0.143 & 28.  \\ \cline{2-9}
        & LOFAR	&94&140&130&27&5.1&49&9600 \\
$7.0$   & MWA	&20&43&43&8.3&2.6&16&3200   \\
        & SKA   &9.1&9.8&8.7&2.0&0.49&2.6&520 \\ 
        & FFTT  &0.59&0.47&0.39&0.098&0.027&0.088&17 \\ \hline	
	
        & Values& 9.9   & 1.3  & 1.6   & 2.3  & 3.1  & 0.58  & 2.   \\	\cline{2-9}
        & LOFAR	&2.2&55&18&73&1.4&5.7&24  \\
$7.5$   & MWA	&4.3&16&4.9&22&1.8&1.8&8.1\\
        & SKA   &0.18&1.7&0.71&2.1&0.076&0.17&0.78 \\   
        & FFTT  &0.0072&0.027&0.015&0.030&0.0023&0.0021&0.012 \\ \hline	
	
        & Values& 2.12  & 1.63 & -0.1  & 1.35 & 1.47 & 0.62 & 0.46 \\ \cline{2-9}
        & LOFAR	&1.6&20&2.1&34&1.2&3.4&6.9 \\
$8.0$   & MWA	&2.7&13&4.2&24&1.5&1.6&2.8 \\
        & SKA   &0.085&0.60&0.090&0.90&0.057&0.095&0.24 \\ 
        & FFTT  &0.0017&0.013&0.0026&0.017&0.0013&0.0014&0.0030 \\
	
\end{tabular}
\end{ruledtabular}
}
\end{table}

\subsubsection{Basic results}

We start by testing assumptions in the ionization power modeling of
$\Pxx$ and $\Pxd$.  
In Table \ref{tab:power} we show the accuracy with which the 21cm
power spectrum can place constraints on the cosmological parameters
from three $z$-bins ranging from $z=6.8 - 8.2$.  We fix the
assumptions concerning $\kmax$, the foreground removal, and the array
layout and specifications, but vary the sophistication with which we
model the ionization power.

Our results agree with those of previous studies
\cite{McQuinn:2005hk,Bowman:2005hj}, \ie 21cm data alone (except for
the optimized FFTT) cannot place constraints comparable with those
from Planck CMB data.  However, if 21cm data are combined with CMB
data, the parameter degeneracies can be broken, yielding stringent
constraints on $\Ok$, $\mnu$ and $\al$.  For example, in the OPT model,
from LOFAR/MWA/SKA/FFTT combined with Planck, the curvature density
$\Ok$ can be measured 5/3/15/78 times better, to a precision
$\Delta\Ok=0.005/0.008/0.002/0.0003$, the neutrino mass $\mnu$ can be constrained
4/9/14/74 times better to accuracy $\Delta\mnu=0.06/0.03/0.02/0.003$, and running
of the scalar spectral index $\al$ can be done 1/2/4/28 times better, to
$\Delta\al=0.002/0.001/0.0006/0.0001$. The more realistic MID model
yields weaker yet still impressive constraints: from SKA/FFTT
combined with Planck, $\Ok$ can be measured 6/109 times better, to
$\Delta\Ok=0.004/0.0002$, $\mnu$ 4/35 times better, to
$\Delta\mnu=0.06/0.007$, and $\al$ 1/15 times better, to
$\Delta\al=0.002/0.0002$. The improved measurements of $\Ok$ and $\al$
enable further precision tests of inflation, 
since generically $\Ok$ is predicted to vanish down to the $10^{-5}$ level, while 
the simplest inflation models (with a single slow-rolling scalar field) 
predict $\al\sim (1-\ns)^2\sim 10^{-3}$. 
For example, the inflaton potential $V(\phi)\propto\phi^2$ predicts 
$\alpha\approx -0.0007$, while $V(\phi)\propto\phi^4$ predicts $\alpha=0.008$.
In addition, 21cm data combined with CMB data
from Planck can make accurate measurements in the mean neutral
fraction $\xH(z)$ at separate redshifts, outlining the full path of
reionization, \eg at the $\Delta\xH(z)\sim 0.01/0.003$ level from
SKA/FFTT data combined with Planck data.

\subsubsection{OPT and MID models}

For most 21cm experiments, the OPT model yields stronger
constraints than the MID model.  The reason is as follows.  By
assuming $\Pxx=\Pxd=0$, there are essentially no neutral fraction
fluctuations in the OPT model.  This means that this model is an ideal
model in which the 21cm power spectrum encodes cosmological
information per se, since $\PDT(\bfk)\propto \sPdd(k)$ at each pixel
in the Fourier space.  In the more realistic MID model, however, the
nuisance ionization parameters has correlations with cosmological
parameters.  Mathematically, the inversion of a correlated matrix
multiplies each error by a degradation factor.

An exception is the FFTT, where the situation is reversed.  As mentioned in Section \ref{sec:asmp-mid}, the sample
variance $\PDT$ in the MID model is smaller than that in the OPT model 
because of two reasons: (i) the MID model assumes non-zero $\Pxx$ and $\Pxd$, and 
$\Pxd$ has negative contribution to the total power spectrum (see Eqs.\ref{eqn:pmu0} and \ref{eqn:pmu2});
(ii) the OPT model assumes $\xH=1$, but $\xH$ takes realistic values (less than 1) in the MID model, decreasing the overall amplitude.  
In a signal-dominated experiment, reduced sample variance can be more important than the degradation from correlations.  

\begin{figure}[ht]
\centering
\includegraphics[width=0.5\textwidth]{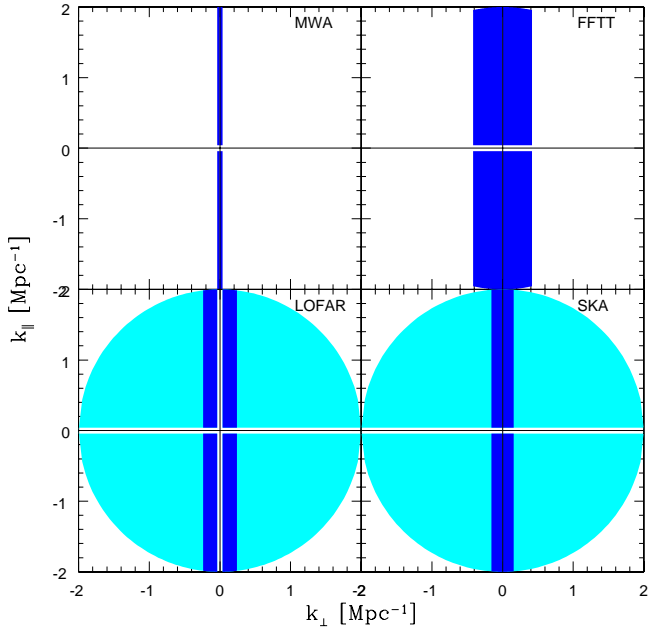} 
\caption{Available $(k_\perp,k_\parallel)$ pixels from MWA (upper left), FFTT (upper right), LOFAR (lower
left) and SKA (lower right), 
evaluated at $z=8$.  The blue/grey regions can be measured with good signal-to-noise from the nucleus and 
core of an array, while the cyan/light-grey regions are measured only with the annulus and have so poor signal-to-noise that they 
hardly contribute to cosmological parameter constraints. 
}
\label{fig:datacube}
\end{figure}

\begin{figure}[h!]
\centering
\includegraphics[width=0.5\textwidth]{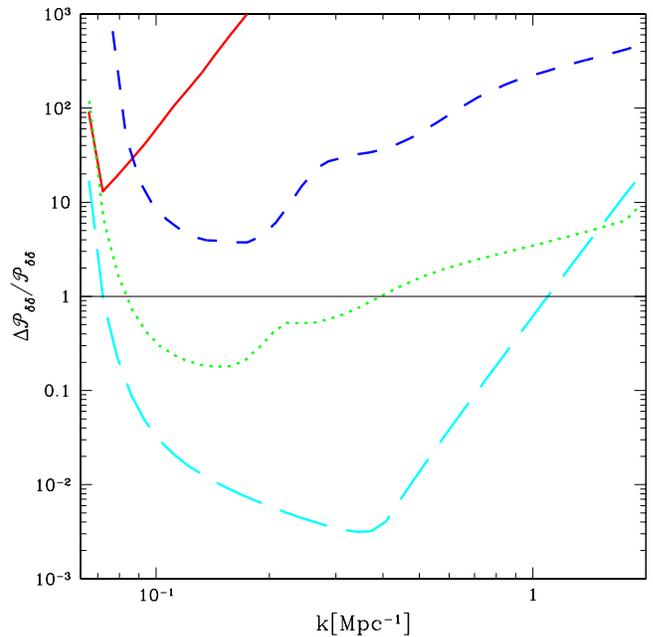} 
\caption{Relative 1$\sigma$ error for measuring $\sPdd(k)$ with the PESS model by observing a 6MHz band that is centered at $z=8$ with MWA (red/solid), LOFAR (blue/short-dashed), SKA (green/dotted) and FFTT (cyan/long-dashed).  The step size is $\Delta \ln k \approx 0.10$.}
\label{fig:Pkerr}
\end{figure}

\subsubsection{PESS model}

Our results show that even combined with CMB data from Planck, the
 21cm data using the PESS model cannot significantly improve constraints.  
There are two reasons for this failure.  Firstly, the PESS model
 essentially uses only $\Pmufour(k)$ to constrain cosmology, by
 marginalizing over $\Pmuzero$ and $\Pmutwo$.  This loses a great deal
 of cosmological information in the contaminated $\Pmuzero$ and
 $\Pmutwo$, in contrast to the situation in the OPT and MID models.
 Secondly, to effectively separate $\Pmufour(k)$ from other two
 moments, the available Fourier pixels should span a large range in
 $\mu$.  Figure \ref{fig:datacube} shows that in MWA and FFTT, the
 data set is a thin cylinder instead of a sphere.  The limitation in
 $\mu$-range will give large degradation factors during the inversion
 of Fisher matrix.  (In the limit that there is
 only one $\mu$ for each shell, then the Fisher matrix is singular and the
 degradation factor is infinite.)  These two factors work together
 with the noise level to shrink the useful $k$-modes into a rather
 narrow range: as shown in Figure \ref{fig:Pkerr}, $\Delta\sPdd <
 \sPdd$ only for $k=0.09-0.4 \; \perMpc$ in SKA, $k=0.07-1 \; \perMpc$ in 
 FFTT and over zero modes in LOFAR and MWA.

\subsection{Varying $\kmax$} \label{sec:kmax}

\begin{figure}[ht]
\centering
\includegraphics[width=0.5\textwidth]{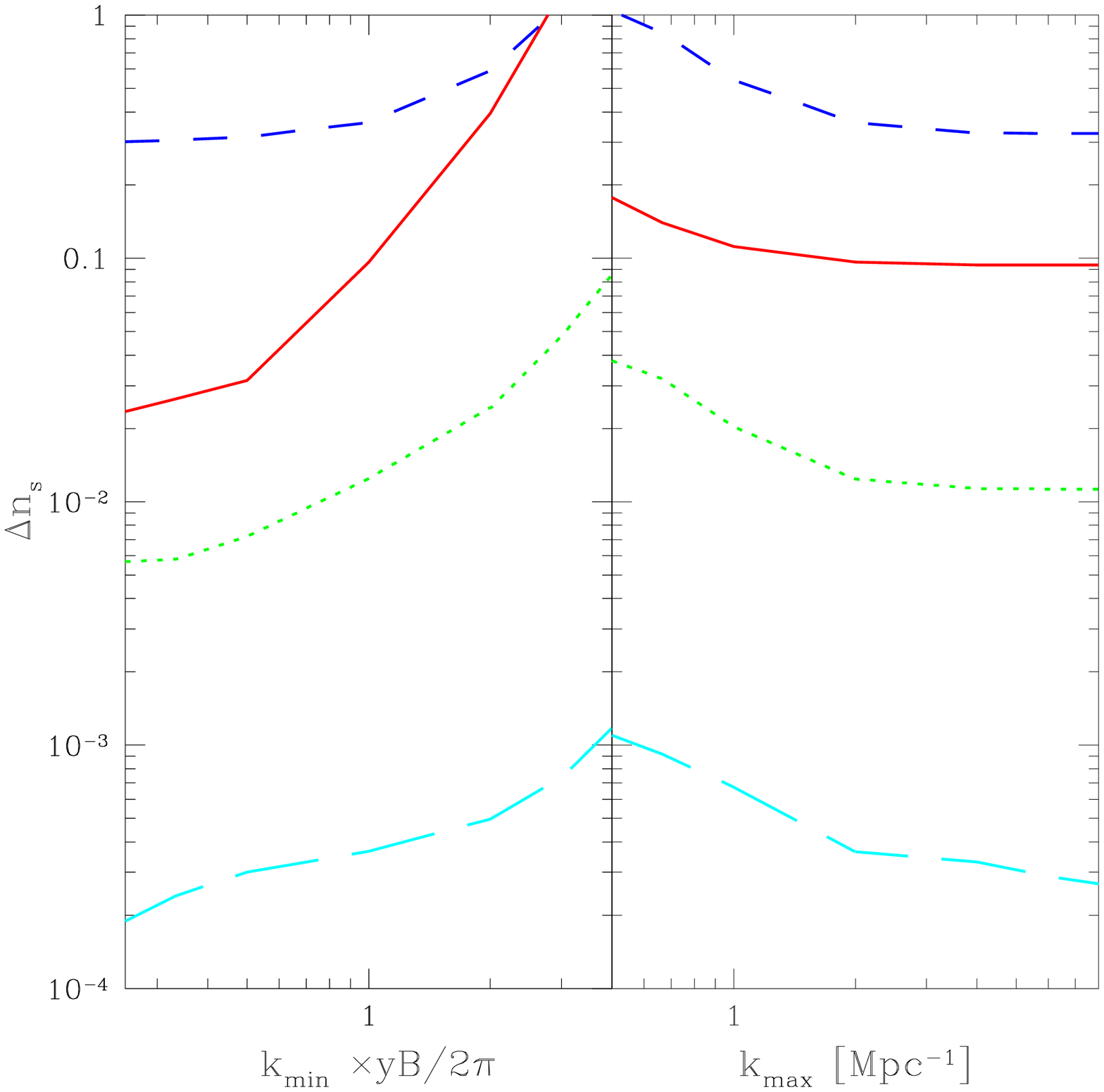} 
\caption{How cosmological constraints $\Delta n_s$ depend on $\kmin$ (left panel) and $\kmax$ (right panel) in the MID model with the 21cm experiments MWA (red/solid), LOFAR (blue/short-dashed), SKA (green/dotted) and FFTT (cyan/long-dashed).  
We plot $\Delta n_s$ in this example because it has the strongest dependence on
$\kmin$ and $\kmax$ of all cosmological parameters.  
The quantity $2\pi/yB$ varies in redshift, so as the horizontal axis of the left panel, we
use the overall scale $\kappa_{\rm min}\equiv \kmin \times (yB/2\pi)$ which is equal for all $z$-bins,    
}
\label{fig:kminkmax}
\end{figure}

We test how varying $\kmax$ affects constraints in this section.  The
cutoff $\kmax$ depends on the scale at which non-linear physics, \eg
the non-linear clustering of density perturbations or the
irregularities of ionized bubbles, enter the power spectrum.  
It is illustrated in the right panel of Figure \ref{fig:kminkmax} that
generically cosmological constraints asymptotically approach a value
as $\kmax$ increases above $\sim 2 \; \perMpc$ (this typical scale can
be larger for cosmic variance limited experiments such as FFTT).  Not
much cosmological information is garnered from these high-$k$ modes
because detector noise becomes increasingly important with $k$. 
The upshot is that the accuracy only weakly depends on $\kmax$.

\subsection{Varying the non-Gaussianity parameter $\xi$}
\label{sec:nongauss_sec}

Table~\ref{tab:omp} shows the effect of changing the non-Gaussianity
parameter $\xi$ in Section~\ref{sec:non-gauss} from the $\xi=1$
(Gaussian) case to $\xi=2$ in the PESS scenario, 
along with changing other assumptions.  
However, there is no need to perform
extensive numerical investigation of the the impact of $\xi$, since it
is readily estimated analytically.  Because $\onesig$ error $\Delta
p_i$ in cosmological parameters is $\sqrt{(F^{-1})_{ii}}$, it follows
directly from \eq{xiDefEq} that $\Delta p$ does not appreciably depend
on $\xi$ for noise dominated experiments like MWA and LOFAR, whereas
$\Delta p \propto \xi^\sigma$ with $\sigma\la 1$ for (nearly) signal
dominated experiments like SKA and FFTT.  Compared with the other
effects that we discuss in this section, this (no more
than linear) dependence on the non-Gaussianity parameter $\xi$ is not
among the most important factors.

\subsection{Varying redshift ranges}
\label{sec:zbin_z}

\begin{table*}
\scriptsize{
\caption{\label{tab:z1} How cosmological constraints depend on the redshift range in OPT model.  Same assumptions as in Table \ref{tab:power} but for different redshift ranges and assume only OPT model.   } 
\begin{ruledtabular}
\begin{tabular}{llccccccccccccc}
    &      & \multicolumn{10}{c}{\it Vanilla Alone} &  & &   \\ \cline{3-12}
		&   z range    & $\Delta\Ol$ & 
 $\Delta\ln(\Omega_m h^2)$   & $\Delta\ln(\Omega_b h^2)$ & 
 $\Delta\ns$		       & $\Delta\ln\As$ & 
 $\Delta\tau$	               & $\Delta\xH(7.0)$ & $\Delta\xH(7.5)$ & 
 $\Delta\xH(8.0)$            & $\Delta\xH(9.2)$            & $\Delta\Ok$ &
 $\Delta\mnu$ [\eV]	       & $\Delta\al$ \\ \hline
 
      & 6.8-10 &0.021&0.20&0.34&0.049&0.67&...&...&...&...&...&0.086&0.75&0.023  \\
LOFAR & 6.8-8.2&0.025&0.27&0.44&0.063&0.89&...&...&...&...&...&0.14&0.87&0.027   \\
      & 7.3-8.2&0.036&0.38&0.61&0.090&1.2 &...&...&...&...&...&0.24&1.3&0.038    \\ \cline{2-15}

      & 6.8-10 &0.037&0.072&0.14&0.016&0.25&...&...&...&...&...&0.031&0.31&0.011 \\
MWA   & 6.8-8.2&0.046&0.11&0.19&0.022&0.37 &...&...&...&...&...&0.056&0.38&0.013 \\
      & 7.3-8.2&0.070&0.15&0.27&0.032&0.51 &...&...&...&...&...&0.097&0.53&0.018 \\ \cline{2-15}
      
      & 6.8-10 &0.0032&0.031&0.061&0.0058&0.12&...&...&...&...&...&0.012&0.096&0.0032  \\
SKA   & 6.8-8.2&0.0038&0.044&0.083&0.0079&0.16&...&...&...&...&...&0.023&0.12&0.0040   \\
      & 7.3-8.2&0.0053&0.059&0.11&0.011&0.21 &...&...&... &...&...&0.042&0.17&0.0054   \\ \cline{2-15}
     
      & 6.8-10 &0.00012&0.0023&0.0058&0.00030&0.011&...&...&...&...&...&0.00045&0.0073&0.00023 \\
FFTT  & 6.8-8.2&0.00015&0.0032&0.0083&0.00040&0.015&...&...&...&...&...&0.00098&0.011&0.00034  \\
      & 7.3-8.2&0.00021&0.0042&0.011&0.00052&0.019 &...&...&...&...&...&0.0021&0.014&0.00043   \\ 
\hline

Planck &     & 0.0070 & 0.0081 & 0.0059 & 0.0033 & 0.0088 & 0.0043 & \nodata & \nodata & \nodata& \nodata & 0.025  & 0.23  & 0.0026 \\ \cline{2-15}
         & 6.8-10 &0.0065&0.0076&0.0057&0.0031&0.0088&0.0043&0.0077&0.0084&0.0082&0.0090&0.0046&0.051&0.0021   \\
\:+LOFAR & 6.8-8.2&0.0066&0.0077&0.0058&0.0031&0.0088&0.0043&0.0077&0.0084&0.0093&...   &0.0051&0.060&0.0022  \\
         & 7.3-8.2&0.0068&0.0079&0.0058&0.0032&0.0088&0.0043&...   &0.0085&0.0093&...   &0.0072&0.081&0.0024   \\ \cline{2-15}

         & 6.8-10 &0.0065&0.0076&0.0056&0.0031&0.0088&0.0043&0.0065&0.0067&0.0066&0.0067&0.0066&0.023&0.0013 \\
\:+MWA   & 6.8-8.2&0.0067&0.0079&0.0057&0.0031&0.0088&0.0043&0.0065&0.0067&0.0069&\nodata&0.0079&0.027&0.0014\\ 	     
         & 7.3-8.2&0.0068&0.0080&0.0058&0.0032&0.0088&0.0043&\nodata&0.0067&0.0069&\nodata&0.011&0.036&0.0017\\ \cline{2-15} 
      
         & 6.8-10 &0.0027&0.0035&0.0045&0.0012&0.0087&0.0042&0.0060&0.0060&0.0060&0.0060&0.0016&0.015&0.00061   \\
\:+SKA   & 6.8-8.2&0.0031&0.0038&0.0046&0.0013&0.0087&0.0042&0.0060&0.0060&0.0060&\nodata&0.0017&0.017&0.00064  \\	       
         & 7.3-8.2&0.0039&0.0047&0.0049&0.0017&0.0087&0.0042&\nodata&0.0060&0.0060&\nodata&0.0020&0.019&0.00075 \\ \cline{2-15}
     
         & 6.8-10 &0.00013&0.0014&0.0033&0.00019&0.0087&0.0042&0.0054&0.0054&0.0054&0.0054&0.00026&0.0025&0.000078 \\
\:+FFTT  & 6.8-8.2&0.00015&0.0015&0.0036&0.00021&0.0087&0.0042&0.0056&0.0056&0.0056&\nodata&0.00032&0.0031&0.000094 \\  
         & 7.3-8.2&0.00020&0.0016&0.0038&0.00023&0.0087&0.0042&\nodata&0.0057&0.0057&\nodata&0.00040&0.0038&0.00011\\ 

\end{tabular}
\end{ruledtabular}
}
\end{table*}

\begin{table*}
\scriptsize{
\caption{\label{tab:z2} How cosmological constraints depend on the redshift range in MID model. Same assumptions as in Table \ref{tab:power} but for different redshift ranges and assume only MID model.}
\begin{ruledtabular}
\begin{tabular}{llccccccccccccc}
    &      & \multicolumn{10}{c}{\it Vanilla Alone} &  & &   \\ \cline{3-12}
		&   z range    & $\Delta\Ol$ & 
 $\Delta\ln(\Omega_m h^2)$   & $\Delta\ln(\Omega_b h^2)$ & 
 $\Delta\ns$		       & $\Delta\ln\As$ & 
 $\Delta\tau$	               & $\Delta\xH(7.0)$ & $\Delta\xH(7.5)$ & 
 $\Delta\xH(8.0)$            & $\Delta\xH(9.2)$            & $\Delta\Ok$ &
 $\Delta\mnu$ [\eV]	       & $\Delta\al$ \\ \hline
 
      & 6.8-10  &0.090&0.055&0.093&0.18&0.43&...&...&...&...&...&0.22&5.7&0.080 \\
LOFAR & 6.8-8.2 &0.13&0.083&0.15&0.36&0.80  &...&...&...&...&...&0.35&12&0.17         \\
      & 7.3-8.2 &0.21&0.099&0.15&0.42&0.81  &...&...&...&...&...&0.62&15&0.18             \\ \cline{2-15}

      & 6.8-10  &0.15&0.012&0.020&0.031&0.46&...&...&...&...&...&0.092&4.4&0.025  \\
MWA   & 6.8-8.2 &0.22&0.017&0.029&0.097&0.76&...&...&...&...&...&0.13&9.6&0.074 	  \\
      & 7.3-8.2 &0.40&0.018&0.030&0.099&1.0&...&...&...&...&...&0.32&18&0.083 		  \\ \cline{2-15}
      
      & 6.8-10  &0.010&0.0031&0.0056&0.0073&0.023&...&...&...&...&...&0.031&0.23&0.0032  \\
SKA   & 6.8-8.2 &0.014&0.0049&0.0081&0.012&0.037&...&...&...&...&...&0.043&0.36&0.0060	 \\
      & 7.3-8.2 &0.018&0.0050&0.0081&0.013&0.039&...&...&...&...&...&0.072&0.41&0.0063  	 \\ \cline{2-15}
     
      & 6.8-10  &0.00029&0.00021&0.00043&0.00025&0.00097&...&...&...&...&...&0.0020&0.0055&0.00011 \\
FFTT  & 6.8-8.2 &0.00041&0.00038&0.00062&0.00036&0.0013&...&...&...&...&...&0.0037&0.0078&0.00017 	   \\
      & 7.3-8.2 &0.00050&0.00039&0.00062&0.00037&0.0013&...&...&...&...&...&0.0058&0.0083&0.00018 	   \\
\hline

Planck &     & 0.0070 & 0.0081 & 0.0059 & 0.0033 & 0.0088 & 0.0043 & \nodata & \nodata & \nodata& \nodata & 0.025  & 0.23  & 0.0026 \\ \cline{2-15}
         & 6.8-10  &0.0069&0.0080&0.0058&0.0032&0.0088&0.0043&0.18&0.26&0.15&0.23&0.017&0.22&0.0026 \\
\:+LOFAR & 6.8-8.2 &0.0070&0.0081&0.0059&0.0032&0.0088&0.0043&0.18&0.26&0.23&...&0.018&0.22&0.0026      \\
         & 7.3-8.2 &0.0070&0.0081&0.0059&0.0032&0.0088&0.0043&...&0.27&0.23&...&0.023&0.22&0.0026 	    \\ \cline{2-15}

         & 6.8-10  &0.0056&0.0065&0.0054&0.0029&0.0087&0.0043&0.32&0.22&0.091&0.36&0.020&0.11&0.0025 \\
\:+MWA   & 6.8-8.2 &0.0061&0.0070&0.0056&0.0030&0.0087&0.0043&0.32&0.22&0.29&...&0.021&0.19&0.0026       \\
         & 7.3-8.2 &0.0061&0.0071&0.0056&0.0030&0.0087&0.0043&...&0.25&0.29&...&0.024&0.19&0.0026 	     \\ \cline{2-15}
      
         & 6.8-10  &0.0025&0.0027&0.0038&0.0023&0.0087&0.0042&0.0094&0.014&0.0075&0.024&0.0032&0.033&0.0020 \\
\:+SKA   & 6.8-8.2 &0.0036&0.0040&0.0044&0.0025&0.0087&0.0043&0.0094&0.014&0.011&...&0.0039&0.056&0.0022	    \\
         & 7.3-8.2 &0.0036&0.0041&0.0044&0.0025&0.0087&0.0043&...&0.015&0.011&...&0.0053&0.056&0.0023		    \\ \cline{2-15}
     
         & 6.8-10  &0.00033&0.00021&0.00043&0.00024&0.0086&0.0042&0.0013&0.0022&0.0030&0.0040&0.00020&0.0052&0.00011 \\
\:+FFTT  & 6.8-8.2 &0.00038&0.00034&0.00059&0.00033&0.0086&0.0042&0.0013&0.0022&0.0031&...&0.00023&0.0066&0.00017 	     \\
         & 7.3-8.2 &0.00041&0.00035&0.00059&0.00033&0.0086&0.0042&...&0.0022&0.0031&...&0.00024&0.0070&0.00017  	     \\

\end{tabular}
\end{ruledtabular}
}
\end{table*}

We now test how accuracies depend on the redshift ranges.  In Table
\ref{tab:z1} (OPT model) and \ref{tab:z2} (MID model), we consider the
optimistic/middle/pessimistic ranges, $z=6.8-10$ / $6.8-8.2$ /
$7.3-8.2$ which is divided by $n_z=4/3/2$ $z$-bins.  The results show
that, from 21cm data alone, the constraints from the extreme ranges
differ significantly (a factor of 5 for $\Delta\Ok$).
Therefore, the sensitivity of a 21cm telescope depends strongly on the
frequency range over which it can observe the signal.

\subsection{Optimal configuration: varying array layout}\label{sec:oc-results}

\begin{figure*}[ht]
\centering
\begin{displaymath}
\begin{array}{ccc} 
  \includegraphics[width=0.33\textwidth]{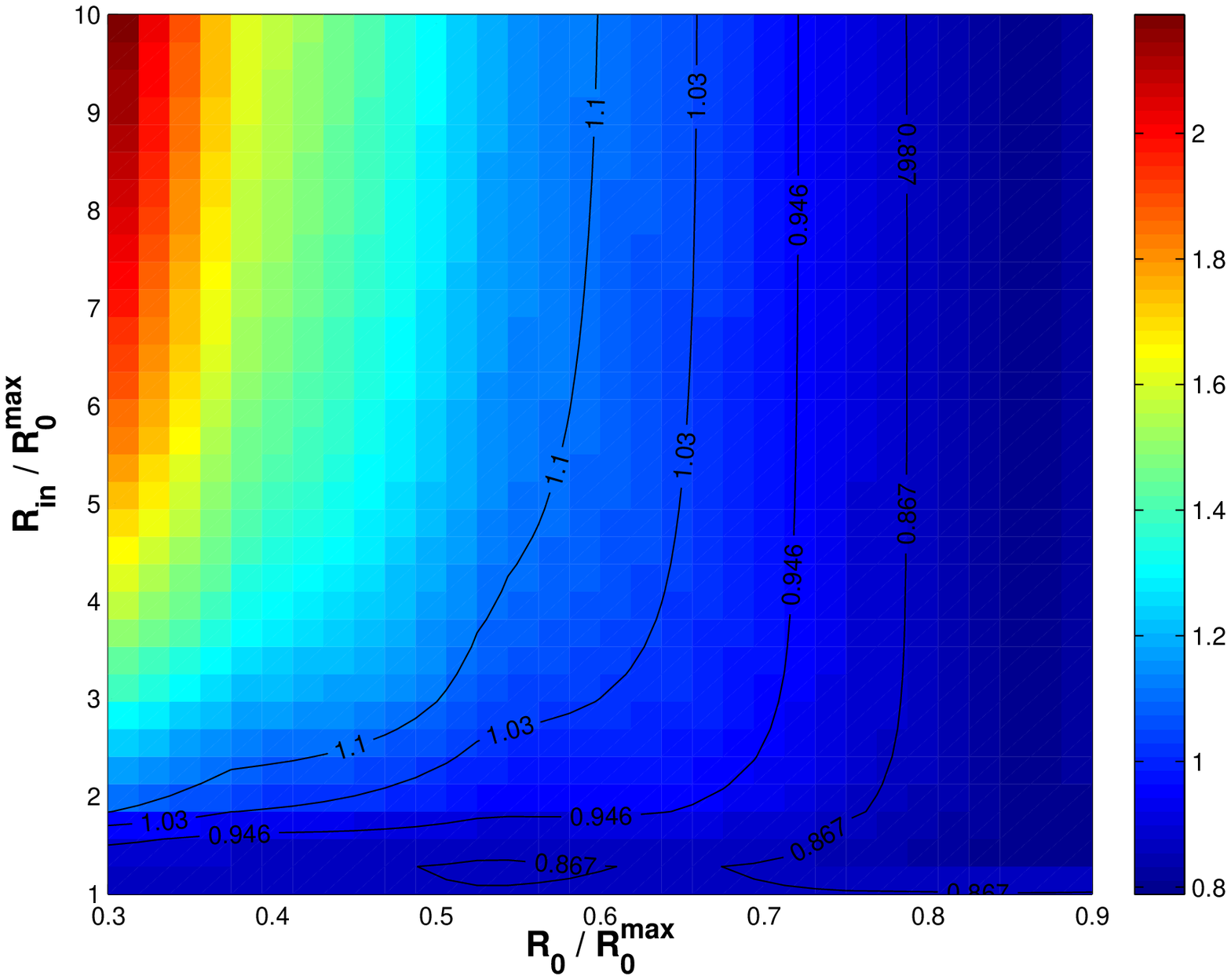} & 
  \includegraphics[width=0.33\textwidth]{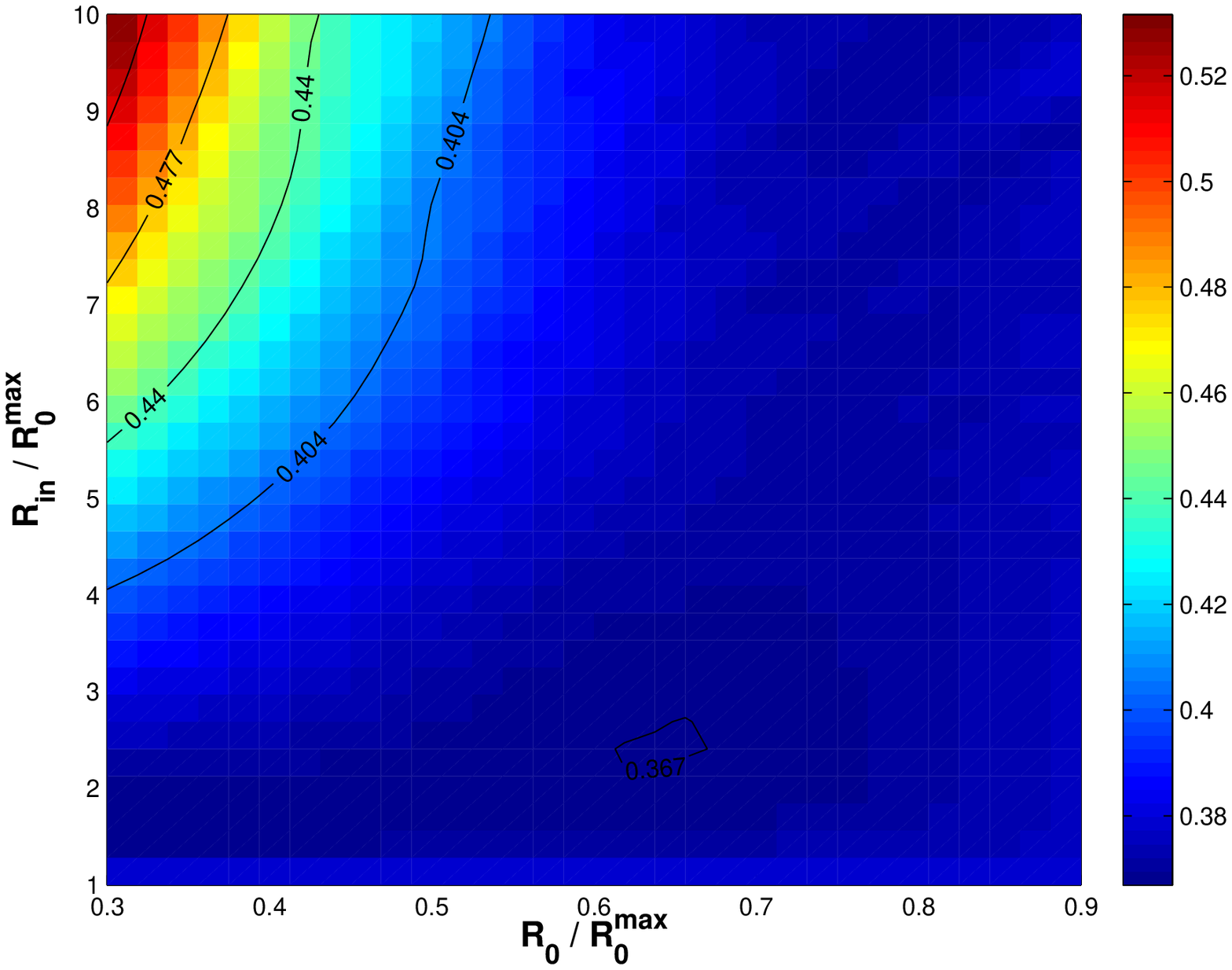} & 
  \includegraphics[width=0.33\textwidth]{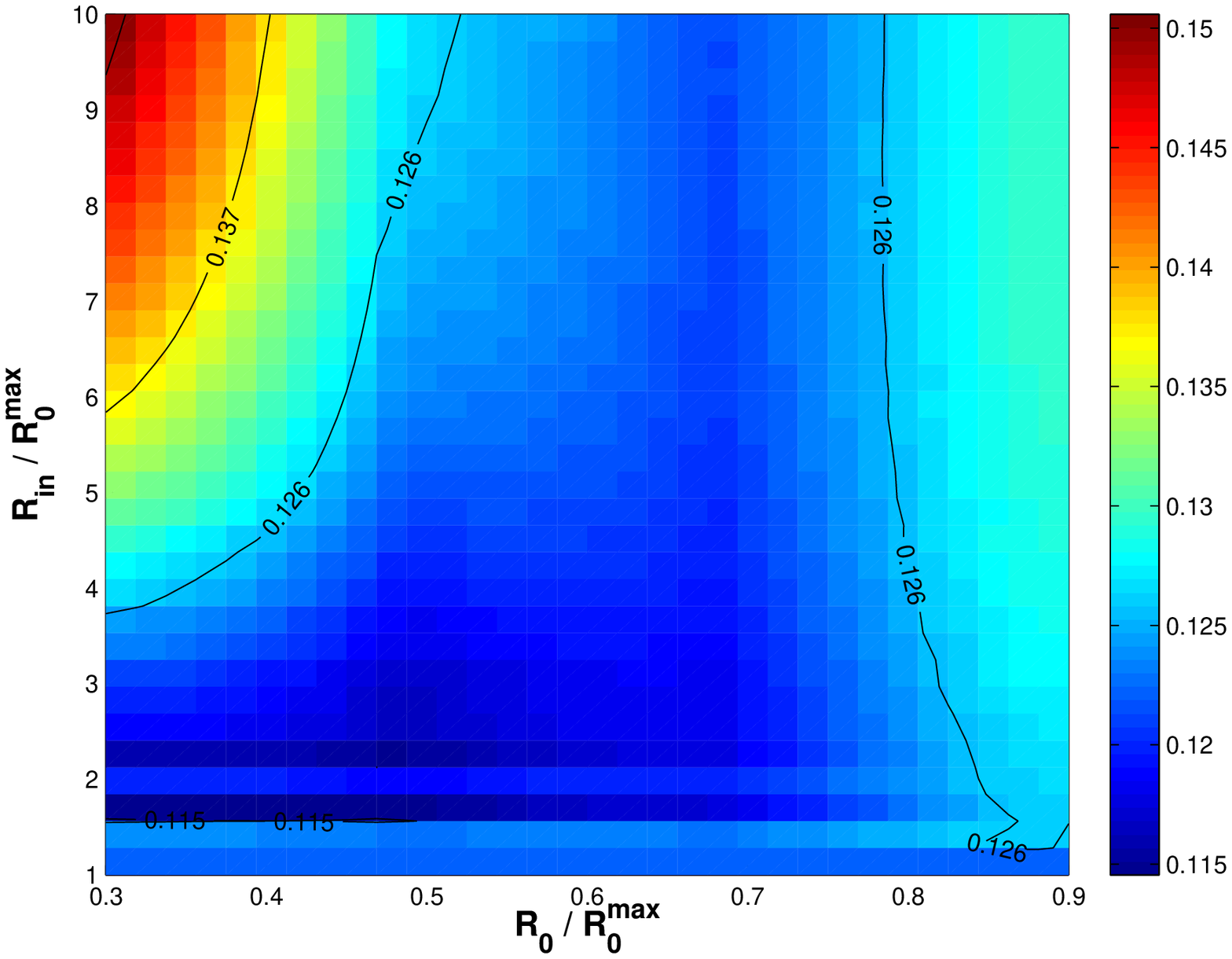} \\
\end{array}
\end{displaymath}
\caption{ 1$\sigma$ error for $\mnu$ marginalized over vanilla parameters for various configuration $(R_0,\, R_{\rm in})$ of LOFAR(left panel), MWA(middle panel) and SKA(right panel).  We made the same assumptions here as in Table \ref{tab:power} but assume only OPT model.}
\label{fig:OC1}
\end{figure*}

\begin{figure*}[ht]
\centering
\begin{displaymath}
\begin{array}{ccc} 
  \includegraphics[width=0.33\textwidth]{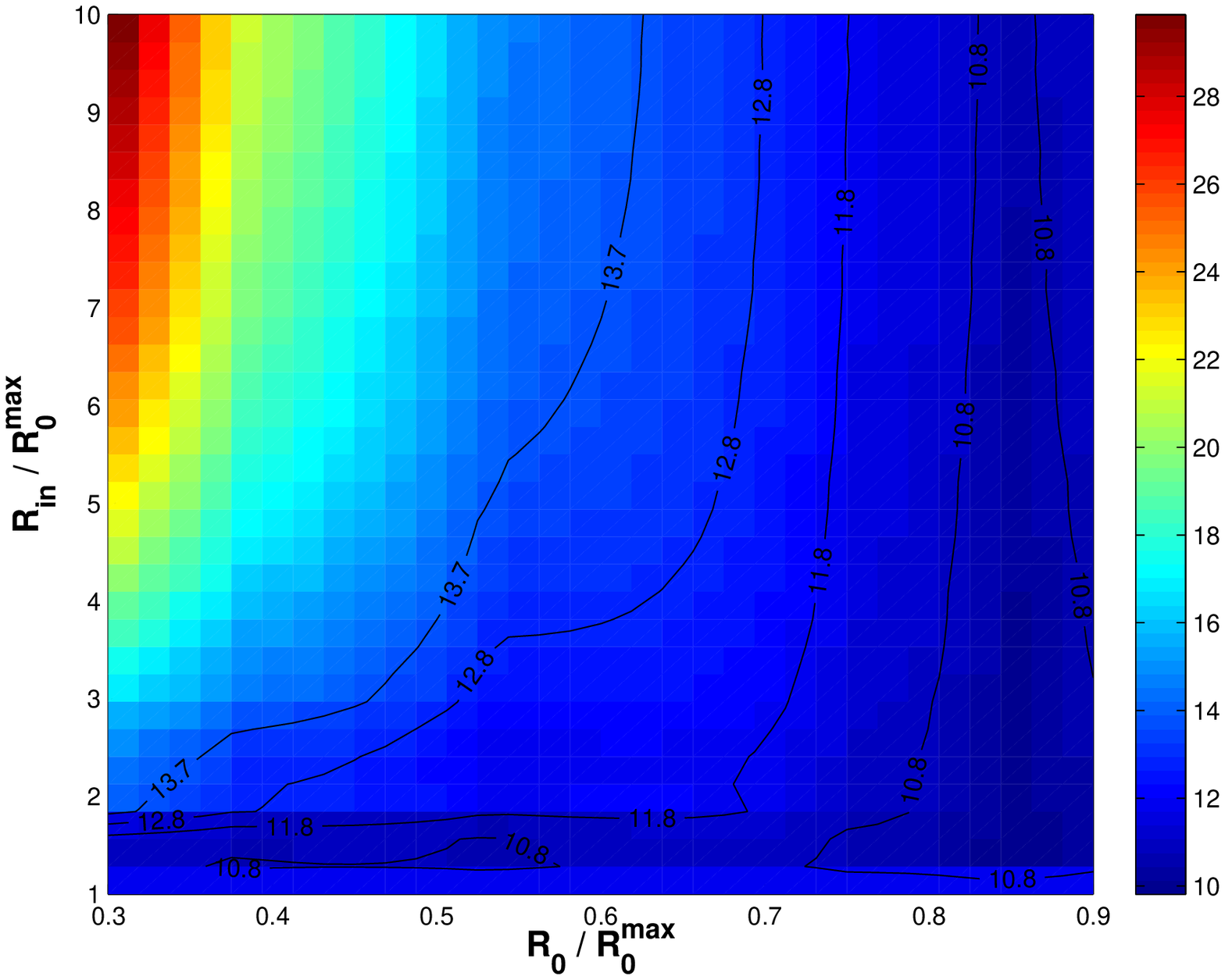} & 
  \includegraphics[width=0.33\textwidth]{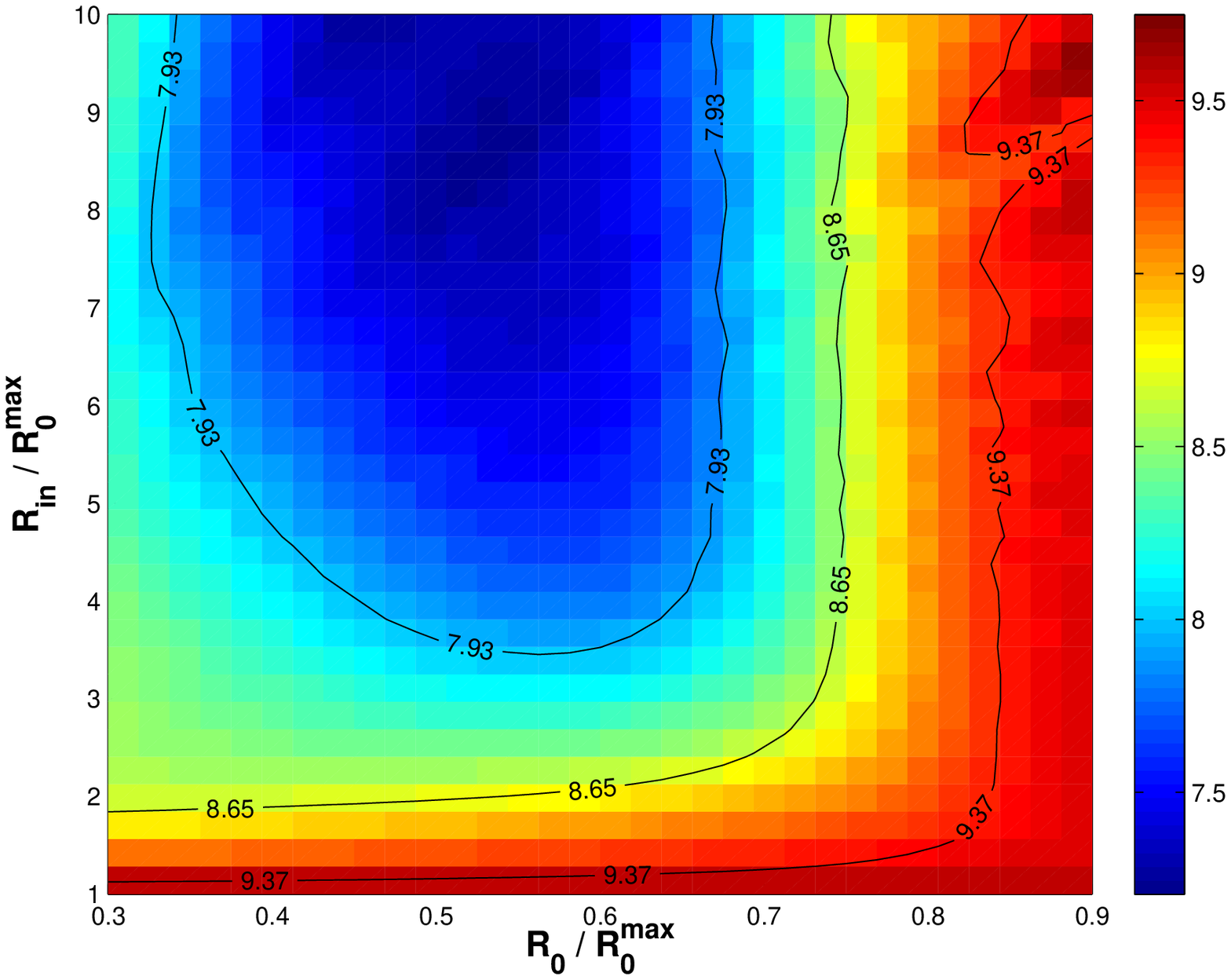} &
  \includegraphics[width=0.33\textwidth]{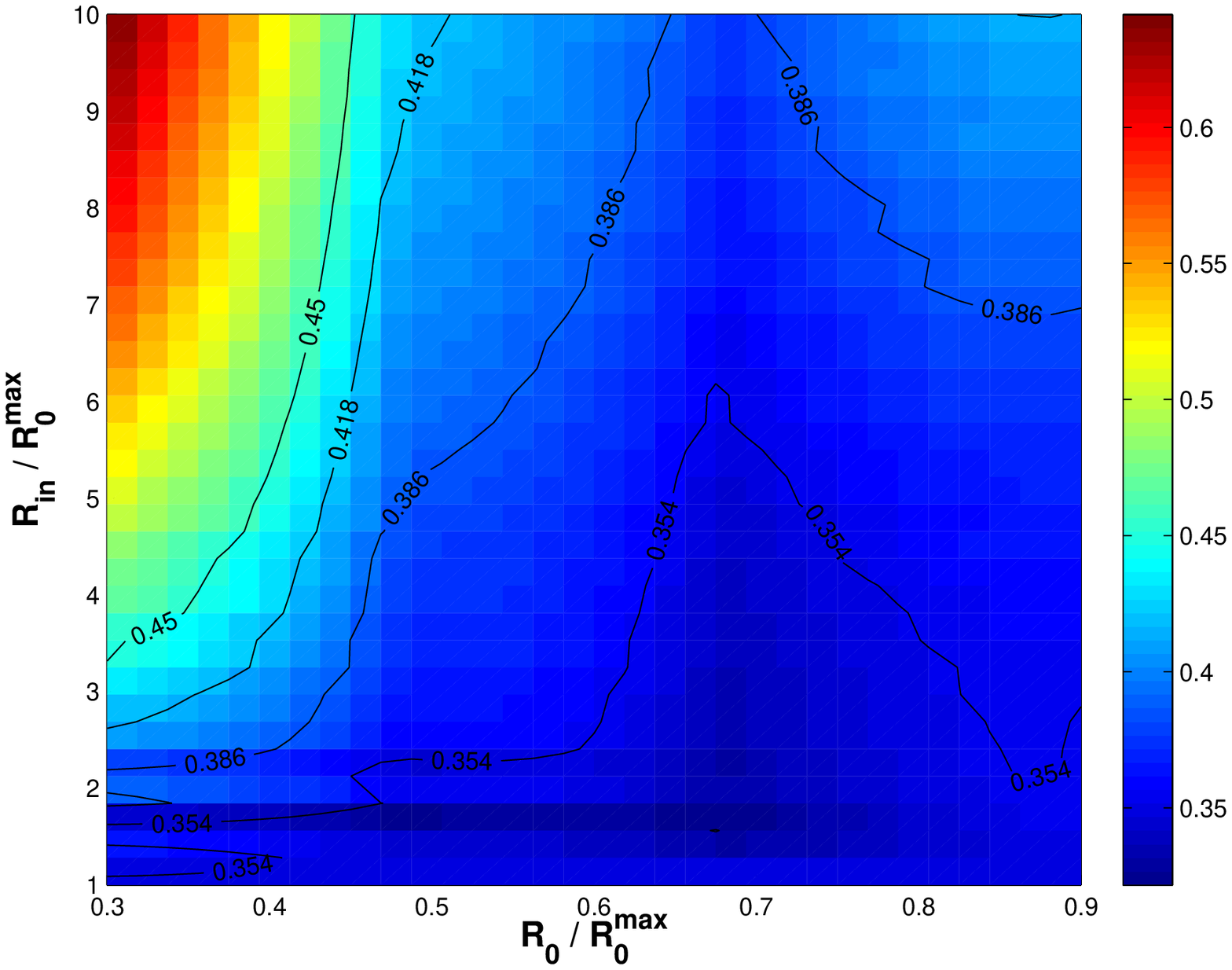} \\
\end{array}
\end{displaymath}
\caption{Same as Figure \ref{fig:OC1} but for MID model.  Figures are for LOFAR(left panel), MWA(middle panel) and SKA(right panel).}
\label{fig:OC2}
\end{figure*}

\begin{table*}
\footnotesize{
\caption{\label{tab:op}  Optimal configuration for various 21cm interferometer arrays.  Same assumptions as in Table \ref{tab:power} but for different array layout.  $R_{\rm in}^{\rm prop}$ is the previously proposed inner core radius.  $\eta$ is the ratio of the number of antennae in the nucleus to the total number inside the core.  $n$ is the fall-off index by which $\rho \propto r^{-n}$ outside the nucleus.}
\begin{ruledtabular}
\begin{tabular}{llccccccl}

                         &    Experiment          &    
$R_0^{\rm max}$ (m) & 
$R_0\,(\times R_0^{\rm max})$	    &	 $R_{\rm in}\,(\times R_0^{\rm max})$	 &     
$R_{\rm in}^{\rm prop}$ (m) \footnotemark[1]   & 
$\eta$         &    $n$ 		& 
Comments      \\   \hline
       & LOFAR   & 319 & 0.84 & 1.28 & 1000 & 0.71 & 6.0 &  Almost a giant core      \\
OPT    & MWA     & 50  & 0.64 & 2.41 & 750  & 0.41 & 3.0 &  Close to a giant core  \\
       & SKA     & 211 & 0.30 & 1.56 & 1000 & 0.09 & 0.83 &  Almost a giant core   \\ \hline
       & LOFAR   & 319 & 0.84 & 1.28 & 1000 & 0.71 & 6.0  &  Almost a giant core \\
MID    & MWA     & 50  & 0.45 & 10   & 750  &  0.20 & 2.3 & {Both a large nucleus and a wide-spread core} \\
       & SKA     & 211 & 0.68 & 1.57 & 1000 & 0.46 & 2.9 &  Almost a giant core \\
\end{tabular}
\end{ruledtabular}
\footnotetext[1]{Note that in LOFAR and SKA there is an annulus with the outer radius 6 km and 5 km respectively.  So for them $R_{\rm in}$ is not the size of total array.}  
}
\end{table*}

In this section we first investigate how array layout affects the
sensitivity to cosmological parameters.  Next, we investigate the
optimal array configuration for fixed antennae number. Our
parametrization of the array configuration is discussed in Section
\ref{configuration}.

We map the constraint in $\mnu$ on the $R_0$--$R_{\rm in}$ plane in
Figure \ref{fig:OC1} (OPT model) and Figure \ref{fig:OC2} (MID model).  $R_0$
is the radius of the compact core, and $R_{\rm in}$ the radius of
inner core, both in the unit of $R_0^{\rm max}\equiv \sqrt{N_{\rm
in}/\rho_0 \pi}$.  Note that if $R_0=R_0^{\rm max}$, then $R_{\rm
in}=R_0^{\rm max}$ --- this is the case of a ``giant
core'', in which all antennae are compactly laid down with a physical
covering fraction close to unity, and is represented by the $x$-axis
in the $R_0$--$R_{\rm in}$ plane (the value of $R_0$ is meaningless if
$R_{\rm in}=R_0^{\rm max}$).  
In Table \ref{tab:op}, we list the
optimal configuration that is indicated by Figure \ref{fig:OC1} and
\ref{fig:OC2}.  The compactness of an array is represented by $R_{\rm
in}/R_0^{\rm max}$, since $R_0^{\rm max}$ is the minimum of $R_{\rm
in}$.  In comparison, $R_0/R_0^{\rm max}$ does not indicate the
compactness, since a slow fall-off configuration with a small $R_0$ is
effectively very close to a giant core.  Rather, $R_0$ is a transition
point from a flat compact core to the fall-off region.
Note that we have three configuration parameters $R_0$, $R_{\rm in}$ and $R_{\rm out}$.  
We find the annulus for SKA and LOFAR to make almost no difference
to the cosmological constraints, 
and therefore focus on how to optimize only the remaining two parameters $R_0$ and $R_{\rm in}$.

Table \ref{tab:op} shows that the optimal layout for OPT model is
close to a giant core, with the inner core much smaller than the
previously proposed.  For MID model, LOFAR and SKA still favors
the quasi-giant-core layout, but MWA favors a large core whose radius
is about the size that was previously proposed.  The accuracies in
$\mnu$ varies in the OPT model by a factor of 3 for LOFAR, 1.4-1.5 for
MWA and SKA, and in the MID model by a factor of 3 for LOFAR, 1.3 for
MWA and 2.2 for SKA.  This means that an optimal
configuration can improve the constraints by a factor up to 3 in noise
dominated experiments, and up to 2 in signal dominated
experiments.

The plots have three interesting features.  First, the configuration
of a quasi-giant core is generically favored.  The reason for this is
that the noise on the temperature in an observing cell with $u_\perp$
is inversely proportional to the square root of the number of
baselines that probe this $u_\perp$.  A compact array increases
the number of baselines that probe small $u_\perp$, reducing the
overall noise level on these modes.  Second, a couple of the upcoming
$21$ cm experiments favor the configuration that is close but not
identical to a giant core.  The reason for this is because arrays become
sample variance limited once they have a certain number of baselines that
probe a given $u_\perp$.  A simple estimate on the signal-to-noise
ratio for a compact MWA shows that on average $\sPdd/\bar{P}_N \approx
5 $ at the $k\sim 0.1 \; \perMpc$ and $\sPdd/\bar{P}_N \approx 1/40$ at
the $k\sim 0.7 \; \perMpc$.  
Although moving more antennae to the center can increase the
signal-to-noise, the error cannot be reduced as much if modes are
already dominated by signal.  Third, in the MID model, MWA favors a
less compact core.  This fact is due to the mixing between
cosmological and ionization parameters.  Remember that the
off-diagonal elements in the Fisher matrix are proportional to the
magnitude of ionization power spectra --- the smaller the magnitude,
the smaller degradation factor and the more accurate is the
cosmological parameter measurement.  Figure \ref{fig:ri} illustrates
that the ionization power spectrum generically falls off at large $k$
such that a relatively large core, which is more sensitive to these
large $k$, may actually improve parameter constraints.  This factor
appears to be important for MWA because, as Figure \ref{fig:datacube}
shows, a compactified MWA only occupies a rather narrow band in
$k$-space.  This means that MWA has to expand significantly in order
to use much more large $k$ modes.

It came to our attention that Lidz \etal \cite{Lidz:2007az} performed
an analysis of the optimal configuration for MWA.  Lidz \etal
\cite{Lidz:2007az} concludes that the optimal layout for MWA is a
giant core.  This conclusion is slight different than ours; we find a
compact but not exactly a giant core is optimal for MWA.  The work in
\cite{Lidz:2007az} defines the optimal configuration to be the
configuration that maximizes the total signal-to-noise, while our
definition is based on parameter constraints. In addition, the
conclusion in \cite{Lidz:2007az} is based on the comparison of a giant
core array configuration to one without a giant core, while we
investigate a range of plausible configurations. It should be pointed
out that both approaches should be tested with detailed simulations.

\subsection{Varying collecting area}

\begin{table}
\footnotesize{
\caption{\label{tab:Ae1}  How cosmological constraints depend on collecting areas in the OPT model.  Same assumptions as in Table \ref{tab:power} but for different collecting areas $A_e$ and assume only OPT model.  The exponent $\beta$ tells the rule of thumb of the $A_e$-dependence of marginalized errors $\Delta p$, assuming $\Delta p\propto (A_e)^\beta$. } 
\begin{ruledtabular}
\begin{tabular}{llccccc}

 	      &  $A_e/A_e^{\rm fid}$ \footnotemark[1]  & 
 $\Delta\Ol$                &$\Delta\ln(\Omega_m h^2)$ & 
 $\Delta\ln(\Omega_b h^2)$  &$\Delta\ns$	    & 
 $\Delta\ln\As$ \\ \hline
         & $ 2.0 \quad$  &0.020&0.24&0.40&0.048&0.80 \\
LOFAR    & $ 1   \quad$  &0.025&0.27&0.44&0.063&0.89 \\
         & $ 0.5 \quad$  &0.039&0.40&0.62&0.10&1.3 \\ \cline{2-7}
	 & $\beta$       & -0.48 & -0.37  & -0.32 & -0.53 & -0.35 \\ \hline
         & $ 2.0 \quad$  &0.057&0.11&0.22&0.021&0.41\\
MWA      & $ 1   \quad$  &0.046&0.11&0.19&0.022&0.37\\
         & $ 0.5 \quad$  &0.042&0.11&0.19&0.027&0.37\\ \cline{2-7}
	 & $\beta$       & 0.22  & 0 & 0.11 & -0.18 & 0.07 \\ \hline
         & $ 2.0 \quad$  &0.0027&0.048&0.099&0.0077&0.19 \\
SKA      & $ 1   \quad$  &0.0038&0.044&0.083&0.0079&0.16 \\
         & $ 0.5 \quad$  &0.0043&0.043&0.076&0.0089&0.15 \\ \cline{2-7}
	 & $\beta$       & -0.34 & 0.08 & 0.19 & -0.10 & 0.17 \\ \hline
         & $ 2.0 \quad$  &0.00014&0.0031&0.0082&0.00037&0.015\\
FFTT   & $ 1   \quad$    &0.00015&0.0032&0.0084&0.00040&0.015\\
         & $ 0.5 \quad$  &0.00017&0.0035&0.0086&0.00046&0.016\\ \cline{2-7}
	 & $\beta$       & -0.14 & -0.09  & -0.03  & -0.16 & -0.05 \\ 
\end{tabular}
\end{ruledtabular}
\footnotetext[1]{$A_e^{\rm fid}$ refers to the fiducial values assumed in Table \ref{tab:spec} and are not the same for different arrays.}
}
\end{table}

\begin{table}
\footnotesize{
\caption{\label{tab:Ae2} How cosmological constraints depend on collecting areas in the MID model.  Same assumptions as in Table \ref{tab:power} but for different collecting areas $A_e$ and assume only MID model.  The exponent $\beta$ tells the rule of thumb of the $A_e$-dependence of marginalized errors $\Delta p$, assuming $\Delta p\propto (A_e)^\beta$.} 
\begin{ruledtabular}
\begin{tabular}{llccccc}

 	      &  $A_e/A_e^{\rm fid}$   & 
 $\Delta\Ol$                &$\Delta\ln(\Omega_m h^2)$ & 
 $\Delta\ln(\Omega_b h^2)$  &$\Delta\ns$	    & 
 $\Delta\ln\As$ \\ \hline
         & $ 2.0 \quad$  &0.086&0.044&0.072&0.15&0.35 \\
LOFAR    & $ 1   \quad$  &0.13&0.083&0.15&0.36&0.80 \\
         & $ 0.5 \quad$  &0.26&0.17&0.35&0.92&2.0 \\ \cline{2-7}
         & $\beta$	 & -0.80 & -0.98 & -1.1 & -1.3 & -1.3 \\ \hline
         & $ 2.0 \quad$  &0.21&0.015&0.025&0.073&0.61\\
MWA      & $ 1   \quad$  &0.22&0.017&0.029&0.097&0.76\\
         & $ 0.5 \quad$  &0.26&0.026&0.045&0.16&1.3\\ \cline{2-7}
	 & $\beta$       & -0.15  & -0.40 & -0.42 & -0.57 & -0.55 \\ \hline
         & $ 2.0 \quad$  &0.013&0.0049&0.0079&0.0092&0.032 \\
SKA      & $ 1   \quad$  &0.014&0.0049&0.0081&0.012&0.037 \\
         & $ 0.5 \quad$  &0.016&0.0063&0.011&0.022&0.053 \\ \cline{2-7}
	 & $\beta$       & -0.15 & -0.18 & -0.24 & -0.63 & -0.36 \\ \hline
         & $ 2.0 \quad$  &0.00036&0.00037&0.00061&0.00032&0.0012\\
FFTT     & $ 1   \quad$  &0.00041&0.00038&0.00062&0.00036&0.0013\\
         & $ 0.5 \quad$  &0.00052&0.00041&0.00066&0.00046&0.0016\\ \cline{2-7}
	 & $\beta$       & -0.27 & -0.07 & -0.06  & -0.26 & -0.21 \\

\end{tabular}
\end{ruledtabular}
}
\end{table}

The survey volume and the noise per pixel are both affected by
changing the collecting area $A_e$ because the solid angle a survey
observes is $\Omega \approx \lambda^2/A_e$ and $P^N \propto 1/A_e^2$
(\Eq{eqn:PNoise}).  For noise-dominated experiments, $\delta \PDT
/\PDT \propto P^N/\sqrt{N_c} \propto A_e^{-2} /
\sqrt{A_e^{-1}}=A_e^{-3/2}$, and, for signal-dominated experiments,
$\delta \PDT /\PDT \propto 1/\sqrt{N_c} \propto A_e^{1/2}$.  If we
parametrize the scaling of the error on a cosmological parameter as
$\Delta p\propto (A_e)^\beta$, we have $-1.5<\beta<0.5$.  A caveat is
FFTT which has fixed $\Omega=2\pi$, so $\delta \PDT /\PDT \propto
A_e^0$ (signal dominated) or $\delta \PDT /\PDT \propto 1/A_e^2$
(noise dominated).  Since FFTT is nearly signal dominated, $\beta\lesssim 0$
for FFTT.

We show how collecting area affects the accuracy in Table
\ref{tab:Ae1} (OPT model) and \ref{tab:Ae2} (MID model).  In the OPT
model, it appears that $\beta\approx -0.4$ for LOFAR, $|\beta|
\lesssim 0.2$ for MWA, $|\beta| \lesssim 0.3$ for SKA, and $\beta\sim
-0.1$ for FFTT.  In the MID model, it appears that $\beta\sim -1.3$
for LOFAR, $\beta\sim -0.5$ for MWA, $\beta\sim -0.6$ for SKA,
$\beta\sim -0.3$ for FFTT.  These exponents are compatible with the
above arguments.  The upshot is that varying $A_e$ does not
significantly affect parameter constraints.

\subsection{Varying observation time and system temperature}

\begin{table}
\footnotesize{
\caption{\label{tab:t0-1}  How cosmological constraints depend on observation time in the OPT model.  Same assumptions as in Table \ref{tab:power} but for different observation time $t_0$ and assume only OPT model.  The exponent $\epsilon$ tells the rule of thumb of the $t_0$-dependence of marginalized errors $\Delta p$, assuming $\Delta p\propto (t_0)^{-\epsilon}$.  $t_0$ is in units of 4000 hours.} 
\begin{ruledtabular}
\begin{tabular}{llccccc}

 	      &  $t_0$   & 
 $\Delta\Ol$                &$\Delta\ln(\Omega_m h^2)$ & 
 $\Delta\ln(\Omega_b h^2)$  &$\Delta\ns$	    & 
 $\Delta\ln\As$ \\ \hline
         & $ 4.0 \quad$  &0.014&0.17&0.28&0.034&0.56\\
LOFAR    & $ 1   \quad$  &0.025&0.27&0.44&0.063&0.89\\
         & $ 0.25\quad$  &0.055&0.56&0.88&0.14&1.8\\ \cline{2-7}
	 & $\epsilon$    & 0.49 & 0.43 & 0.41 & 0.51 & 0.42 \\ \hline
         & $ 4.0 \quad$  &0.040&0.081&0.16&0.015&0.29\\			  
MWA      & $ 1   \quad$  &0.046&0.11&0.19&0.022&0.37\\			  
         & $ 0.25\quad$  &0.059&0.15&0.27&0.038&0.52\\ \cline{2-7} 	  
	 & $\epsilon$	 & 0.14 & 0.22 & 0.19 & 0.34 & 0.21 \\ \hline 
         & $ 4.0 \quad$  &0.0019&0.034&0.070&0.0054&0.13\\			  
SKA      & $ 1   \quad$  &0.0038&0.044&0.083&0.0079&0.16\\			  
         & $ 0.25\quad$  &0.0060&0.061&0.11&0.013&0.21\\ \cline{2-7} 	  
	 & $\epsilon$	 & 0.41 & 0.21 & 0.16 & 0.32 & 0.17 \\ \hline 
         & $ 4.0 \quad$  &0.00014&0.0031&0.0082&0.00037&0.015\\			  
FFTT     & $ 1   \quad$  &0.00015&0.0032&0.0084&0.00040&0.015\\			  
         & $ 0.25\quad$  &0.00017&0.0035&0.0086&0.00046&0.016\\ \cline{2-7} 	  
	 & $\epsilon$	 & 0.07 & 0.04 & 0.02 & 0.08 & 0.02 \\ 

\end{tabular}
\end{ruledtabular}
}
\end{table}

\begin{table}
\footnotesize{
\caption{\label{tab:t0-2}  How cosmological constraints depend on observation time in the MID model. Same assumptions as in Table \ref{tab:power} but for different observation time $t_0$ and assume only MID model. The exponent $\epsilon$ tells the rule of thumb of the $t_0$-dependence of marginalized errors $\Delta p$, assuming $\Delta p\propto (t_0)^{-\epsilon}$.  $t_0$ is in units of 4000 hours. } 
\begin{ruledtabular}
\begin{tabular}{llccccc}

 	      &  $t_0$   & 
 $\Delta\Ol$                &$\Delta\ln(\Omega_m h^2)$ & 
 $\Delta\ln(\Omega_b h^2)$  &$\Delta\ns$	    & 
 $\Delta\ln\As$ \\ \hline
         & $ 4.0 \quad$  &0.061&0.031&0.051&0.11&0.25\\
LOFAR    & $ 1   \quad$  &0.13&0.083&0.15&0.36&0.80\\
         & $ 0.25\quad$  &0.36&0.24&0.50&1.3&2.9\\ \cline{2-7}
	 & $\epsilon$    & 0.64 & 0.74 & 0.82 & 0.89 & 0.88 \\ \hline
         & $ 4.0 \quad$  &0.15&0.010&0.017&0.052&0.43\\			  
MWA      & $ 1   \quad$  &0.22&0.017&0.029&0.097&0.76\\			  
         & $ 0.25\quad$  &0.36&0.037&0.064&0.23&1.8\\ \cline{2-7} 	  
	 & $\epsilon$	 & 0.32 & 0.47 & 0.48 & 0.54 & 0.52 \\ \hline 
         & $ 4.0 \quad$  &0.0089&0.0035&0.0056&0.0065&0.022\\			  
SKA      & $ 1   \quad$  &0.014&0.0049&0.0081&0.012&0.037\\			  
         & $ 0.25\quad$  &0.023&0.0090&0.015&0.031&0.075\\ \cline{2-7} 	  
	 & $\epsilon$	 & 0.34 & 0.34 & 0.36 & 0.56 & 0.44 \\ \hline 
         & $ 4.0 \quad$  &0.00036&0.00037&0.00061&0.00032&0.0012\\			  
FFTT     & $ 1   \quad$  &0.00041&0.00038&0.00062&0.00036&0.0013\\			  
         & $ 0.25\quad$  &0.00052&0.00041&0.00066&0.00046&0.0016\\ \cline{2-7} 	  
	 & $\epsilon$	 & 0.13 & 0.04 & 0.03 & 0.13 & 0.10 \\ 

\end{tabular}
\end{ruledtabular}
}
\end{table}

The detector noise is affected by changing the observation time and system temperature.  From \Eq{eqn:PNoise}, the noise $P^N \propto T_{\rm sys}^2/t_0$.  Therefore, for noise dominated experiments, $\delta \PDT /\PDT \propto P^N/\sqrt{N_c} \propto T_{\rm sys}^2/t_0$, and for signal dominated experiments, $\delta \PDT /\PDT \propto 1/\sqrt{N_c} \propto (T_{\rm sys}^2/t_0)^0$.  Assuming that errors in cosmological parameter $\Delta p \propto (T_{\rm sys}^2/t_0)^\epsilon$, we have $0<\epsilon<1$.  

Since $T_{\rm sys}^2$ and $t_0^{-1}$ shares the same exponent, we
evaluate the $\epsilon$ by varying only $t_0$ in Table \ref{tab:t0-1}
(OPT model) and \ref{tab:t0-2} (MID model).  It appears that in
average $\epsilon\sim 0.5$ for LOFAR, $\epsilon\sim 0.3$ for MWA,
$\epsilon\sim 0.3$ for SKA, $\epsilon < 0.1$ for FFTT in the OPT
model, and $\epsilon\sim 0.8$ for LOFAR, $\epsilon\sim 0.5$ for MWA,
$\epsilon\sim 0.4$ for SKA, $\epsilon \lesssim 0.1$ for FFTT in the
MID model.  These exponents are compatible with the expected
$0<\epsilon<1$ from the above argument.  The upshot is that the order
unity changes in $T_{\rm sys}$ and $t_0$ play a marginal role in the
accuracy for future signal-dominated experiments.

\subsection{Varying foreground cutoff scale $\kmin$}

Finally, we test how accuracy is affected by varying $\kmin$ 
above which foregrounds can be cleaned from the signal.  
One expect that the constraints tend to approach 
asymptotic values at small enough $\kmin$.  However, 
the most effectively constrained modes are at
small $k$ ($k \sim 0.1 ~{\rm Mpc}^{-1}$) for noise dominated
experiments, while the contributions from larger $k$ modes are more
important for cosmic variance limited experiments.  This means that $\kmin$ 
affects the noise dominated experiments most.  
Left panel of Figure \ref{fig:kminkmax} illustrates this by plotting 
cosmological constraints as a function of the relative minimum cutoff 
$\kappa_{\rm min} \equiv \kmin \times y(z)B(z)/2\pi$ which is a constant scale 
factor for all $z$-bins by definition.  The slopes at 
$\kappa_{\rm min} = 1$ are rather large for MWA (varying from $\kappa_{\rm min} = 0.5$ 
to $2$, the error in $n_s$ varies from 0.032 to 0.39, about 10 times larger).  
For a signal dominated experiment like SKA, the constraints can be off by a factor of 3, 
or FFTT by a factor of 1.6.  This suggests that in general $\kmin$ is among top factors to affect cosmological constraints.

\section{Conclusion \& outlook}\label{sec:conclusion}

\subsection{Which assumptions matter most?}

\begin{figure}[ht]
\centering
\includegraphics[width=0.5\textwidth]{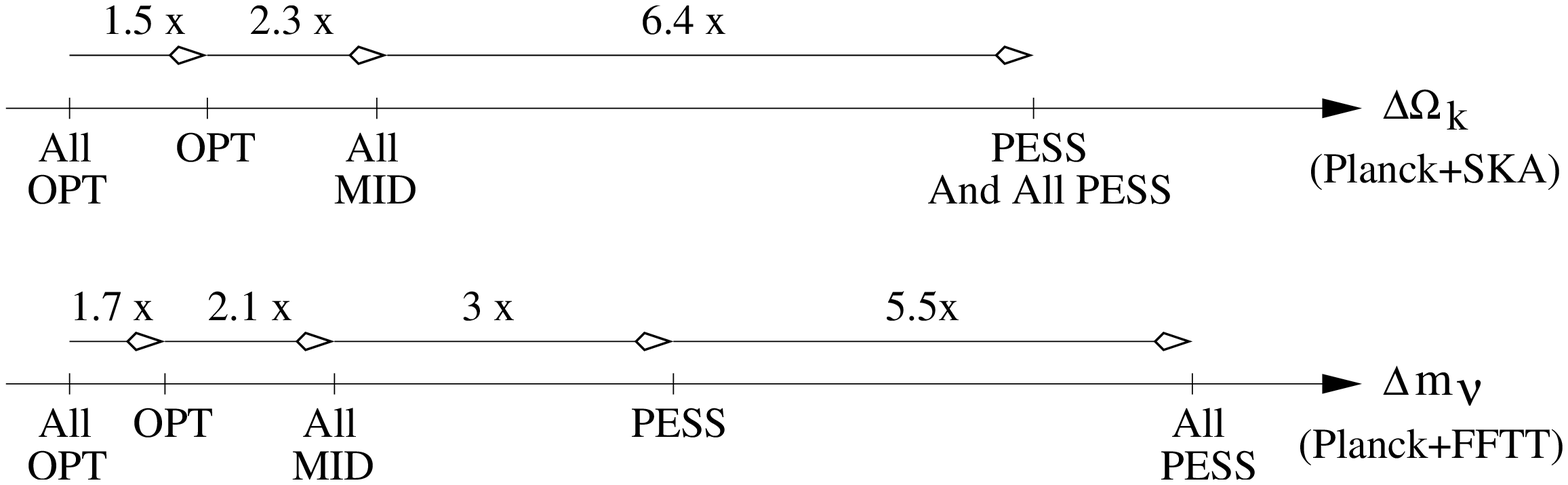} 
\caption{Cartoon showing how cosmological parameter measurement accuracy depends on various assumptions. The cases labeled merely ``PESS'' or ``OPT'' 
have the PESS/OPT ionization power spectrum modeling with MID assumptions for everything else.
}
\label{fig:lineup}
\end{figure}

In Section \ref{sec:results}, we have quantified how cosmological
parameter measurement accuracy depends on assumptions about ionization
power modeling, reionization history, redshift range, experimental
specifications such as the array configuration, and astrophysical
foregrounds.  We now return to the overarching question from
Section~\ref{sec:intro} that motivated our study: among these
assumptions, which make the most and least difference?

To quantify this, we consider two of the parameters for which 21cm
tomography has the most potential for improving on Planck CMB
constraints based on our estimates: $\Ok$ and $\mnu$.  Figure
\ref{fig:lineup} shows $\Delta\Ok$ based on data from Planck plus SKA
as well as $\Delta\mnu$ from Planck plus FFTT.  Varying
the ionization power modeling from PESS to OPT models improves the
constraints on these two parameters by a factor of 6--15.
From 21cm data alone in the OPT model, the optimal array configuration
can affect accuracies up to a factor 3 (Figure~\ref{fig:OC1}),
redshift ranges affect it by up to a factor of 5 (Table~\ref{tab:z1}),
and residual foregrounds affect it by up to a factor of 10
(Figure~\ref{fig:kminkmax}, left panel).  In summary, the assumptions
can be crudely ordered by importance as ionization power modeling
$\gg$ foregrounds $\sim$ redshift ranges $\sim$ array layout $ > A_e
\sim T_{\rm sys} \sim t_0 \sim\kmax \sim$ non-Gaussianity.

\subsection{Outlook}

We have investigated how the measurement of cosmological parameters
from 21 cm tomography depends on various assumptions.  We have found
that the assumptions about how well the reionization process can be
modeled are the most important, followed by the assumptions pertaining
to array layout, IGM evolution, and foreground removal.

Our results motivate further theoretical and experimental work. On the
theoretical side, it will be valuable to develop improved EoR data
analysis techniques.  The OPT approach is restricted to when neutral
fraction fluctuations are not important, which is not an accurate
approximation during the EOR.  On the other hand, although the PESS
approach is in principle insensitive to our poor understanding of
reionization by marginalizing over it, in practice this approach
destroys too large a fraction of the cosmological information to be
useful.  Hopefully more detailed EoR simulations will
enable our MID approach to be further improved into a phenomenological
parametrization of our ignorance that is robust enough to be reliable,
yet minimizes the loss of cosmological information.
\footnote{
It is also possible to constrain cosmological parameters using lensing of 21cm 
fluctuations \cite{Zahn:2005ap,Metcalf:2006ji,Metcalf:2008gq,Zhang:2006fc}.}

On the experimental side, there are numerous complications that are
beyond the scope of this paper, but that are important enough to
deserve detailed investigation in future work.  To what extent can
radio-frequency interference be mitigated, and to what extent does it
degrade cosmological parameter accuracy? This is particularly
important for instruments in densely populated parts of the world,
such as LOFAR.  To what extent is the subtraction of the foreground
point sources hampered by the complicated off-center frequency scaling of
the synthesized beam?  To what extent does the dramatic variation of
the synchrotron brightness temperature across the sky affect our
results and optimal array design?  Performing a realistic end-to-end
simulation of possible experiments (from sky signal to volts and back)
should be able to settle all of these issues.

These are difficult questions, but worthwhile because the potential
for probing fundamental physics with 21 cm tomography is impressive: a
future square kilometer array optimized for 21 cm tomography could
improve the sensitivity of the Planck CMB satellite to spatial
curvature and neutrino masses by up to two orders of magnitude, to
$\Delta\Omega_k\approx 0.0002$ and $\Delta m_\nu\approx 0.007$ eV, and
detect at $4\sigma$ the running of the spectral index predicted by the
simplest inflation models.

\begin{acknowledgments}
The authors wish to thank Judd Bowman, Jacqueline Hewitt and Miguel Morales 
for helpful discussions and comments.  
YM thanks Yi Zheng for technical help.  
This work is supported by 
the U.S.~Department of Energy (D.O.E.) under cooperative research agreement DE-FC02-94ER40818, 
NASA grants NAG5-11099 and NNG06GC55G, NSF grants AST-0134999 and 0607597, and from the David and Lucile
Packard Foundation and the Research Corporation.  OZ was supported by the Berkeley Center for Cosmological Physics.
\end{acknowledgments}

\appendix

\section{$\chi^2$ goodness of fit in the MID model}
\label{chi2}

In this appendix, we elucidate some issues in separating cosmological information from astrophysics in the MID model, and give the $\chi^2$ goodness-of-fit test.

The parametrization of ionization power spectra is based on the assumption that these power spectra are smooth functions of $k$, and therefore can be parametrized with as many parameters as necessary to fit the data at some accuracy.  However, the separation of cosmology from astrophysics implicitly depends on another assumption that the shapes of ionization power spectra are distinguishable from that of matter power spectrum, since one can only measure the \emph{total} 21cm power spectrum.  Albeit sometimes the shape may be similar at small $k$ (see the plateaus in the ratios of power spectra in Figure \ref{fig:ri}), the slope and amplitude of ionization power spectrum at the fall-off region can in principle distinguish nuisance functions from the matter power spectrum, determine the overall amplitude, and in return use the data at small $k$ to further constrain the nuisance parameters that correspond to the amplitudes.  

An avalanche of data from upcoming 21cm experiments can make it possible to justify the MID model with some parametrization of ionization power spectra.  
There are standard statistical methods for testing whether the parametrization is successful.  We now give a compact description of the $\chi^2$ goodness-of-fit test, and refer interested readers to \cite{PDG} for a useful review on the statistics.  
Note that we did not implement the $\chi^2$ test in this paper since this would need observational data.  The description of $\chi^2$ test below is intended to complement the discussions of the MID model in the main part of this paper.  
We want to test the hypothesis $H_0$ that the parametrization with fitting parameter values is an accurate account of the ionization power spectra.  The parameter vector to be fitted is $\Theta\equiv \left( \lambda_i\,(i=1,\ldots,N_p),\beta_\al\,(\al=1,\ldots,n_{\rm ion})\right)$, where $N_p$ and $n_{\rm ion}$ are the number of cosmological and ionization parameters, respectively.  The observed data vector is $\bfy\equiv (y_1,\ldots,y_N)$ where $y_i \equiv \PDT(\bfk_i)$ at each pixel $\bfk_i$ labeled by $i=1,\ldots, N$, where $N$ is the total number of pixels.  Assuming the Gaussian statistic in the measurements, the corresponding vector $\bfF$ for the expected value is  
$F(\bfk_i;\Theta)=(\sPdd-2\sPxd+\sPxx)+2(\sPdd-\sPxd)\mu^2+\sPdd\mu^4$, and the variance is $ \sigma_i^2 \equiv (\delta \PDT(\bfk_i))^2 = \frac{1}{N_c}[\PDT(\bfk_i)+P_N(k_{i\,\perp})]^2$.  
We can now compute $\chi^2$: 
\beq{eq:chi2}
\chi^2(\Theta)=(\bfy-\bfF(\Theta))^T C^{-1} (\bfy-\bfF(\Theta))\,, 
\een 
where $C$ is the covariance matrix.  If each measurement $y_i$ is independent, then $C$ becomes diagonal with $C_{ii}=\sigma_i^2$.  Then \Eq{eq:chi2} is simplified to be 
\ben
\chi^2(\Theta)=\sum_{i=1}^{N} \frac{ [y_i - F(\bfk_i;\Theta)]^2}{\sigma_i^2}\,.
\een
We can define the $p$-value as the probability, under the assumption of the hypothesis $H_0$, of obtaining data at least as incompatible with $H_0$ as the data actually observed.  So 
\ben
p = \int_{\chi^2(\Theta)}^{\infty} f(z;n_d) dz\,,
\een
where $f(z;n_d)$ is the $\chi^2$ probability density function (p.d.f.) with $n_d$ degrees of freedom $n_d=N-(N_p+n_{\rm ion})$.  Values of the $\chi^2$ p.d.f. can be obtained from the CERNLIB routine PROB \cite{prob}. 
To set the criterion, a fit is good if $p \ge 0.95$, \ie the real data fit the parametrization better than the 95\% confidence level.

\end{document}